\documentclass{sig-alternate-10pt}

\usepackage{pslatex}
\usepackage{balance}
\usepackage{xcolor}
\usepackage{url}
\usepackage[margin=1in]{geometry}

\newcommand{\spara}[1]{\smallskip\noindent{\bf #1}}

\newcommand{\eat}[1]{}

\begin{document}
\pagestyle{empty}

\title{Graph Data Management and Graph Machine Learning: Synergies and Opportunities}

\author{Arijit Khan$^1$ \,\, Xiangyu Ke$^2$ \,\, Yinghui Wu$^3$  \\
\affaddr{$^1$Aalborg University, Denmark \,\, $^2$Zhejiang University, China \,\, $^3$Case Western Reserve University, USA}\\
\email{$^1$arijitk@cs.aau.dk \,\, $^2$xiangyu.ke@zju.edu.cn \,\, $^3$yxw1650@case.edu}
}

\maketitle

\vspace{-1in}

\begin{abstract}
The ubiquity of machine learning, particularly deep learning, applied to graphs is evident in applications ranging from cheminformatics (drug discovery) and bioinformatics (protein interaction prediction) to knowledge graph-based query answering, fraud detection, and social network analysis.
Concurrently, graph data management deals with the research and development of effective, efficient, scalable, robust, and user-friendly systems and algorithms for storing, processing, and analyzing vast quantities of heterogeneous and complex graph data.
Our survey provides a comprehensive overview of the synergies between graph data management and graph machine learning, illustrating how they intertwine and mutually reinforce each other across the entire spectrum of the graph data science and machine learning pipeline. Specifically, the survey highlights two crucial aspects:
{\bf (1)} How graph data management enhances graph machine learning, including contributions such as improved graph neural network performance through graph data cleaning, scalable graph embedding, efficient graph-based vector data management, robust graph neural networks, user-friendly explainability methods; and
{\bf (2)} how graph machine learning, in turn, aids in graph data management, with a focus on applications like query answering over knowledge graphs and various data science tasks.
We discuss pertinent open problems and delineate crucial research directions.
\end{abstract}

\section{Introduction}
\label{sec:introduction}

\medskip
\medskip

Graph data, ranging from social and biological networks to financial transactions, knowledge bases, and transportation systems,
permeates various domains. In these graphs, nodes represent entities with distinct features, while edges capture relationships between them.
The growing volume of graph data and the increasing demand to extract value in real applications necessitate effective graph data management (GDM).
Broadly speaking, data management encompasses a suite of algorithms and systems for acquiring, validating, storing, organizing, protecting, and processing data so they can be easily found and queried effectively, efficiently, securely, and cost-effectively.
The principle of data management is to optimize data usage and comply with regulations, so to enable fair and responsible decision making, while maximizing the utility in downstream tasks.
Modern data management challenges include the three V's of big data (volume, velocity, and veracity), dirty data, secure and distributed data processing, cloud computing, usability, new data types, emerging applications, etc.
While general data management focuses on handling structured or semi-structured data such as tables and logs, graph data management presents unique challenges due to the interconnected nature of graph data.
Managing relationships, traversals, and graph-specific queries (e.g., communities or reachabilities) demand specialized algorithms and data structures. Additionally, the irregularity and scale of graphs introduce challenges in indexing, storage, and real-time updates that go beyond traditional DM solutions.
Specialized graph database management systems (graph DBMS), e.g., Neo4j, TigerGraph, Microsoft Cosmos DB, and Amazon Neptune were developed supporting graph transactions, queries, visualization, and diverse data models \cite{Tian22}.

Machine learning (ML), a subfield of artificial intelligence (AI), uses algorithms to learn knowledge from data and generalize to unseen cases, often without explicit programming.
Key principles of ML include data representation, performance evaluation on downstream tasks, and iterative optimization to improve accuracy.
Based on the above requirements, ML models and systems are developed and deployed ensuring that they are effective, efficient, robust, and user-friendly.
As ML becomes mainstream, there is a growing focus on explainability, transparency, fairness, safety, trust,  and ethical decision-making.
Graph machine learning (GML), in particular, graph neural networks (GNNs) have shown great promises for graph data-centric applications, such as classification, link prediction, community detection, question answering, and recommendation \cite{wu2020comprehensive}.

\begin{figure}[t!]
 \centering
 \includegraphics[scale=0.43]{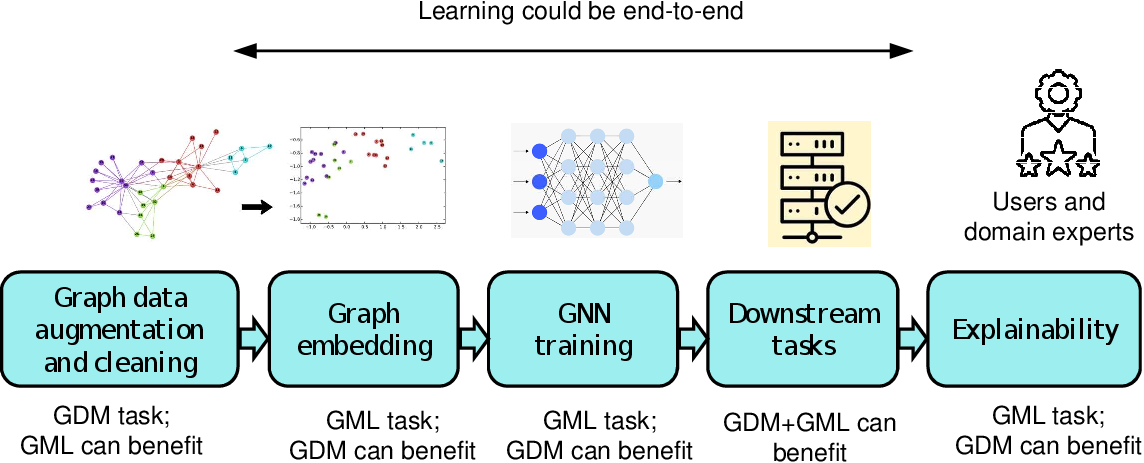}
 \caption{Graph data pipeline in data science and machine learning applications. Graph embedding can be task-specific or task-agnostic. Graph neural network (GNN) training can be end-to-end based on downstream tasks. We show which phases belong to GDM and which belong to GML, and can benefit from each other.}
 \label{fig:pipeline}
 \vspace{-5mm}
\end{figure}

While data management (DM) and machine learning (ML) serve distinct purposes, their synergy is essential, as data is foundational to both.
{\bf First,} effective collaboration between data management and ML is necessary to unleash the full potential of an organization's data. For instance, data management techniques ensure clean, reliable, and up-to-date datasets, enabling ML models to generate accurate and trustworthy insights.
{\bf Second}, in modern data science applications, complex data undergo various processes involved in machine learning to generate the final predictive output, collectively forming a data pipeline \cite{Polyzotis0WZ17, Kumar0017}.
Figure~\ref{fig:pipeline} illustrates a representative graph data pipeline, encompassing the early stages of the graph data extraction, integration, cleaning, acquisition, validation, and enrichment;
intermediate stages dealing with graph embedding, vector data, graph neural network (GNN) training, AutoML;
and concluding stages involving downstream tasks and human-in-the-loop interactions, such as explaining the results of black-box GNN models.
Managing effective and efficient data pipelines increases the need for robust data management solutions.
{\bf Third}, ML approaches enhance DM functionalities, e.g., ML can automate data transformation processes and might also understand a user's query intent to improve querying performance.
Recent graph systems with ML capabilities \cite{abs-2303-14617,HorchidanC23,
Abdallah023} highlight the need to explore the synergies between two related fields: GDM and GML.
Emerging technology landscapes such as AI, ML, edge computing, serverless and cloud computing, modern hardware, Internet-of-Things, data lakes, and Large Language Models (LLMs) are expanding the domain of data-driven downstream applications and what is feasible including real-time decision-making capabilities, streamlining integration, and enhanced security, making the synergy even more critical.

This survey examines the interplay between GDM and GML across different stages of the data pipeline depicted in Figure~\ref{fig:pipeline}.
We identify three key scenarios to structure the survey:
{\bf (a)} when GDM benefits GML; {\bf (b)} when GML enhances GDM; and finally {\bf (c)} when GDM + GML integration facilitates downstream tasks.
For example, the initial phase of graph data cleaning is a GDM task, where we explore GML's contributions (\S \ref{sec:cleaning}).
In contrast, stages like graph embedding, GNN training, and explainability focus on GML objectives, with GDM systems improving their efficiency and effectiveness (\S \ref{sec:embedding}, \S \ref{sec:index}, and \S \ref{sec:explainability}).
The fourth phase about downstream tasks benefits significantly from the synergy of GDM and GML, as discussed in \S \ref{sec:gml4gdm}.

\spara{Motivation: What are new in GDM and GML?}
With the rapid advances of graph machine learning (GML) techniques, such as graph embedding \cite{00010YAWP021}, GNNs \cite{WuPCLZY21,ZhangCZ22,ma2021deep}, graph transformers \cite{abs-2202-08455}, graphGPT \cite{abs-2310-13023}, foundation models \cite{abs-2310-11829}, and LLMs for graphs \cite{jin2023large, abs-2311-12399}, the role of graph data management (GDM) in the GML lifecycle has become increasingly vital.
This spans all stages of the data pipeline, including preparation, improvement, embedding, training, and explanation.
Recently, both academia and industry have emphasized the need for high-quality, large-scale data and robust, scalable, secure, and explainable models in ML systems \cite{Polyzotis0WZ17,abs-2309-10979}.
While there exist surveys and tutorials
discussing the synergy between data management and ML -- primarily focusing on
relational data and relational database management systems \cite{ChaiWLNL23,
Kumar0017,Polyzotis0WZ17}, similar resources about graph data are comparatively scarce.
Both GDM and GML pose significant challenges as follows.

In terms of GDM, graph data are inherently irregular, with nodes and edges forming complex, variable-length connections. This contrasts with the strict schema of relational data, where rows and columns provide a predictable structure.
Therefore, specialized GDM systems are often required to quickly navigate and retrieve complex multi-hop neighbors.
This places unique demands on GDM for efficient sampling and traversal strategies.

Due to their interconnected nature, partitioning graph nodes without disrupting critical structural properties, such as community boundaries, poses significant challenges.
Unlike traditional data partitioning, where rows in tables can often be divided without impacting data relationships, graph partitioning must preserve inter-node dependencies to maintain the graph's integrity, typically requiring extensive communication between nodes or servers.
GDM systems have to manage this inter-partition communication efficiently to support applications at scale -- a challenge that is usually less pronounced in relational data management where tables can often be processed more independently.

Last but not least, graphs can be both high-dimensional and sparse, especially real-world graphs containing billions of nodes, but relatively few edges per node. Storing and efficiently retrieving meaningful patterns from these sparse yet high-dimensional graphs cause difficulties that do not typically arise with dense tabular data.

Analogously, GML poses significant challenges due to the non-IID and unnormalized nature of graph data, the absence of strict schema, and irregular structures.
Unlike traditional ML, where data samples are often processed independently, graph data require interdependent computations, leading to increased computational costs.
Optimizing GML systems for model training, such as by supporting distributed training with efficient data loading and caching, is essential but challenging.

GML also generates high-dimensional embeddings for nodes and edges, which are crucial for tasks like node classification, link prediction, and similarity search. Managing these embeddings is more demanding than handling traditional ML data with simpler numeric or categorical features. Efficient storage, indexing, and retrieval mechanisms, such as vector databases or hybrid storage solutions, are essential for managing the large-volume high-dimensional embeddings produced by GML.

Additionally, reasoning in GML relies heavily on combining intricate feature interactions with graph topology.
Graph data often require advanced feature engineering based on structural motifs or specific subgraph patterns.
This demands robust support for pattern matching and subgraph extraction within GML systems, capabilities that are rarely needed in tabular ML.
Scaling these operations for large graphs is particularly challenging and requires effective indexing and optimization strategies.

Finally, heterogeneity and multimodal graph data (e.g., graphs with nodes and edges having text and image-based features), along with emerging applications beyond classification and prediction (e.g., entity resolution \cite{Li0SZAW20}, knowledge graphs question-answering \cite{YasunagaRBLL21}, graph combinatorial optimizations \cite{VelickovicYPHB20}), and ``black-box'' deep learning approaches introduce further complexities to the deployment of GNNs.

Against this backdrop, our survey
covering a set of the latest solutions that integrate GDM and GML techniques is both timely and relevant.
We believe that our survey will attract and promote interdisciplinary research
that advances
scalable and explainable data pipelines for new data challenges in graph analysis.

\spara{Roadmap.}
In this survey, we demonstrate how graph data management and machine learning facilitate each other at different stages
in a graph data pipeline. In particular, we delve into the following topics:

\noindent$\bullet$ Benefits of graph data cleaning and augmentation in improving the GNN performance (\S \ref{sec:cleaning});

\noindent$\bullet$ Application of graph data management algorithms and systems for scalable graph embedding learning (\S \ref{sec:embedding});

\noindent$\bullet$ Vector data management using graph-based indexes (\S \ref{sec:index});

\noindent$\bullet$ GNN explainability methods, focusing on their usability and robustness (\S \ref{sec:explainability});

\noindent$\bullet$ Application of graph machine learning in knowledge graphs query answering (\S \ref{sec:kg}); and

\noindent$\bullet$  Applications of graph-based retrieval augmented generation (graph RAG) in large language models (LLMs) for data science tasks (\S \ref{sec:graphrag}).

We discuss background and related work in \S \ref{sec:background} and \S \ref{sec:related}, respectively, and conclude with future work in \S \ref{sec:conclusions}.

\section{Background}
\label{sec:background}

\medskip
\medskip

We introduce background materials on graph neural networks and graph embeddings.

{\bf Graph neural networks (GNNs)} are deep learning models to tackle graph-related
tasks in an end-to-end manner \cite{WuPCLZY21}. GNNs have many variants, e.g.,
graph convolutional network (GCN) \cite{KipfW17}, graph attention network (GAT) \cite{VelickovicCCRLB18},
graph isomorphism network (GIN) \cite{XuHLJ19}, GraphSAGE \cite{HamiltonYL17}, graph auto-encoder \cite{KW16},
graph generative adversarial network (GraphGAN) \cite{WangWWZZZXG18}, and APPNP \cite{KlicperaBG19}, etc.
Specifically, graph convolution operations can be categorized as spectral \cite{BrunaZSL13}
and spatial \cite{DuvenaudMABHAA15}
approaches. In spectral methods, filters are applied on a graph's frequency modes computed via graph Fourier transform.
Spectral formulations rely on the fixed spectrum of the graph Laplacian, and are suitable only for graphs with a single
structure (and varying features on nodes), as well as are computationally expensive. On the other hand, spatial methods
are not restricted to a fixed graph structure, as they extract local information by propagating features between neighboring nodes.
Kipf and Welling \cite{KipfW17} also develop a first-order approximation of the spectral convolution, which results
in propagation between neighboring nodes. In particular, GCN
adopts a general form as follows.
\begin{equation}
\label{eq-prop}
    X^k = \delta(\Hat{D}^{-\frac{1}{2}} \Hat{A} \Hat{D}^{-\frac{1}{2}} X^{k-1}  \Theta^{k})
\end{equation}

Here $\Hat{A} = A + I$, where $I$ represents the identity matrix and $A$ is the adjacency matrix of graph $G$. $X^k$ indicates node feature representation in the $k$-th GCN layer, (with $X^0=X$ a matrix of input node features). $\Hat{D}$ represents the diagonal node degree matrix of $\Hat{A}$, $\delta(.)$ is the non-linear activation function, and $\Theta^{k}$ represents the learnable weight matrix for the $k$-th layer.
State-of-the-art GNNs follow a similar feature learning paradigm: Update the features of every node by aggregating the counterparts from its neighbors.
The inference cost of feature propagation-based GNNs is usually polynomial-time \cite{ChenWDL00W20,KlicperaBG19}.
GNNs have been employed in node and graph classification (e.g., GCN \cite{KipfW17}, GAT \cite{VelickovicCCRLB18},
GraphSAGE \cite{HamiltonYL17}, GIN \cite{XuHLJ19}), link prediction (e.g., LGLP \cite{CaiLWJ22}), and entity resolution
(e.g., GraphER \cite{Li0SZAW20}), etc.

{\bf Graph embedding} or {\bf representation learning} \cite{CaiZC18,CuiWPZ19} generates low-dimensional
representation vectors of nodes, edges, and graphs that capture the structure and features of graphs accurately for downstream ML tasks.
Graph embedding algorithms can be categorized into three classes. {\bf (1)} {\em Matrix factorization} methods \cite{QiuDMLWT18}
construct feature representations based on
the adjacency or Laplacian matrix, and exploit spectral techniques.
{\bf (2)} {\em Random-walk} methods \cite{GroverL16} transform a graph into a set of random walks via sampling and then employ
Skip-Gram to generate embeddings.
{\bf (3)} {\em Graph neural networks} (GNNs)-based approaches \cite{HamiltonYL17,VelickovicCCRLB18} focus on generalizing graph spectra into semi-supervised or
supervised graph learning. They often follow a recursive neighborhood aggregation
scheme to generate embeddings. State-of-the-art matrix factorization and random walk methods
generally work on homogeneous graphs where nodes and edges share the same type, and the algorithms
consider only graph structures. In contrast, GNN-based approaches exploit both graph structures
and node features. They can be end-to-end, implying that the learning of embeddings is
implicit within the GNN model and computed in a task-dependent manner. Embeddings of more complex networks such as
heterogeneous information networks \cite{Sun00CXWY18}, relational graphs \cite{SchlichtkrullKB18}, hypergraphs \cite{ACPSSY23},
knowledge graphs \cite{AliBHVGSFTL22}, uncertain graphs \cite{HuCHFL17}, signed networks \cite{YuanWX17},
dynamic graphs \cite{BMVZ21}, spatio-temporal networks \cite{SiGXYDP22} have been studied.

\section{Graph Data Management \\ for Graph ML}
\label{sec:gdm4gml}

\medskip
\medskip

We discuss applications of graph data management such as data cleaning and augmentation in improving the GNN performance, graph algorithms, databases, and systems for scalable embedding learning, graph indexes for vector data management, and graph view-based explanation generation to enhance usability.

\subsection{Graph Data Cleaning and \\ Augmentation}
\label{sec:cleaning}

\medskip
\medskip

Enhancing graph data to improve
the performance of graph learning has seen an increased interest \cite{abs-2309-10979}.
Existing data augmentation techniques from computer vision and natural language processing research cannot be easily generalized to irregular-shaped graph data.
Graph data augmentation (GDA) \cite{abs-2202-08871} specifies enriching graph data to improve graph learning, which is categorized into
``editing-based'' and ``representation-based''.
Editing-based methods aim to
derive graph editing operations,
such as removal, addition, or modify
nodes, edges, features, or (sub)graphs
~\cite{0003LNW0S21,ZhaoZW21,ZhangZSKK22,HanJLH22,gdet}, to improve the model performance
such as graph neural networks.
These methods may follow a deterministic
process, learning to derive editing operations, or
via a stochastic editing process.
Graph sparsification \cite{ZhengZCSNYC020}, condensation \cite{JinZZLTS22}, and diffusion \cite{ZhaoDDKT21} are also
applied to improve GNN-based analysis.

Instead of deriving graph
editing operators, representation based GDA
directly learns to refine graph representation
to improve follow-up analytical tasks.
These methods train learnable parameters to generate augmented samples or graph representation
and may adopt
structure learning, adversarial training, contrastive learning, or automated augmentation \cite{0001XYLWW21,ZhaoTZJ0SASYJ22,SureshLHN21}.
Compared with representation-based approaches, editing-based
GPA may be more
interpretable and explainable, by performing data provenance
analysis over the derived editing operators.
On the other hand,
representation-based approaches
can be readily streamlined
as input for downstream
(graph) learning tasks,
hence, may serve better
in the need of
end-to-end learning pipelines.

Error detection and repairing
have been studied for graphs using rules and logic-based
solutions, such as graph dependencies \cite{FanL19} and graph keys \cite{FanFTD15}, neighborhood constraints \cite{JBM24,linrepairing},
uncertain edges cleaning \cite{LinPCX17}.
Graph association rules (GARs) \cite{FanJLLTZ20}
detect missing links and semantic errors in graphs, while assisting in link prediction.
GNNCleaner \cite{XLXTWLL23} repairs node labels to improve GNN robustness against label noise.
Recent work such as SHACTOR~\cite{rabbani2023shactor}
extracts validated shapes (a graph pattern carrying
value, topological or cardinality constraints) with configurable
measurements such as support to
detect anomalies in knowledge graphs for
error detection and cleaning.
In general,
rule-based error detection treats and
deals with each error scenario
in an isolated manner and
often falls short of capturing
complex scenarios
where errors are from multiple
sources with different forms,
and may require additional effort to
be adapted for general error detection.

Graph learning
has been introduced to improve
error detection and repairing for graphs.
 Generative adversarial learning and
active learning has been
exploited to improve
error detection in
graphs~\cite{gdet, guan2023gale}. For example, GALE~\cite{guan2023gale}
supports an interactive
active, generative adversarial detection framework
for graph error detection.
The method applies few-shot learning
to learn an error generation model
that best fits a limited number of
examples of different types of errors, and applies the
model to augment
the detected errors via a generative
adversarial model to detect more
errors. Active learning is adopted in
this process to assist the
error generation in the GAN-based
error detection.

\spara{Synergy.} There are good  opportunities to integrate and interact with machine learning and graph data cleaning towards ML-based
graph data cleaning systems.
(1) Graph data constraints and
rules can be exploited to characterize
domain knowledge and context for
ML data cleaning models.
These graph data constraints and
rules also provide a validation mechanism to make ML-based data cleaning reasonable.
For example, graph association rules~\cite{FanJLLTZ20} or
validation shapes~\cite{rabbani2023shactor}
can be equipped with
learnable domain-specific patterns to
improve the quality of domain-specific knowledge graphs.
(2) The domain knowledge, context, and data constraints may also be properly featurized for potential
training of foundational data cleaning
models. The expressive ML models can be fine-tuned to perform downstream data cleaning
tasks without conducting isolated,
from-scratch data cleaning pipelines.

On the other hand,
 learning for
graph error detection still requires
a properly large amount of
high-quality annotated
examples, which remain a luxury for many applications such as
domain sciences. Scaling
ML solutions to large-scale
graph cleaning also calls for efficient graph
learning algorithms.
Moreover, making ML-empowered
data cleaning explainable
with domain knowledge
remains desirable yet a missing
feature in current data systems.
These provide opportunities for
emerging needs such as
fact checking tools
in scientific knowledge graphs.

\subsection{Scalable Graph Embedding \\ and GNN Training}
\label{sec:embedding}

\medskip
\medskip

The surge of billion-scale graphs emphasizes the importance of efficient embedding learning on large graphs, as well as GNN training with them,
such as for link prediction on Twitter with over one billion edges \cite{GuptaGLSWZ13}, users and products recommendation at Alibaba \cite{WangHZZZL18}, etc.
To scale GNNs to large
graphs, various sampling strategies, e.g.,  node-wise sampling, layer-wise sampling, and graph-wise sampling are adopted \cite{LinYYFPCX23}.

To resolve efficiency and scalability issues with large graphs, recent works mainly focus on parallel computation,
distributed systems, CPU-GPU hybrid architecture, and  new hardware. PANE enables scalable and attributed networks
embedding by measuring node attribute affinity with random walks, embedding computation via joint matrix factorization,
and using multi-core parallelization \cite{YangSX0LB20}. DistGER exploits information oriented distributed random walks
and distributed Skip-Gram learning for scalable graph embedding \cite{FangKLWFLYC23}.
GraphVite \cite{ZhuXTQ19} employs a CPU-GPU hybrid architecture, simultaneously performing graph random walks on
CPUs and embedding training on GPUs. Marius \cite{MohoneyWXRV21} optimizes data movements between CPU and GPU on a
single machine for large KG embedding. Seastar \cite{WuMCJLZCY21} develops a novel GNN training
framework on GPUs with a vertex-centric programming paradigm.
XGNN \cite{TWCWYZL24} designs a multi-GPU GNN training system to fully utilize GPU and CPU memory and high-speed interconnects.
Amazon released DistDGL \cite{Zheng0WZSSGZK20},
a distributed graph embedding framework
with mini-batch training using the Deep Graph Library (DGL).
Facebook's Pytorch Biggraph \cite{LererWSLWBP19} exploits graph partitioning and parameter servers to learn large-graph
embeddings on multiple CPUs using PyTorch. ReGNN develops ReRAM-based  architecture for GNN acceleration \cite{LLJLZDXL22}.

\spara{Synergy.}
Both graph embedding and GNN training are GML tasks. We showcase how GDM techniques can enhance them in four major ways: algorithms and systems, software-hardware co-design, and graph databases.

\noindent $\bullet$ {\em Efficient algorithms}. To improve efficiency and scalability of GNN training often at the cost of accuracy loss, mini-batch training and sampling strategies are developed, which can scale with data parallelism. Parallel and distributed training algorithms aim at reducing computation and communication overheads and design effective graph partitioning methods, all of which deal with irregularity, inter-connectedness, and sparseness in graph structure. Random walks approximate GNN message passing (e.g., APPNP \cite{KlicperaBG19}) and capture neighborhood structures for generating graph embedding. Therefore, improving the effectiveness of random walks, reducing their numbers and path lengths, as well as distributed random walk mechanisms have great potentials to improve the efficiency and scalability of GNN training and graph embedding. Efficient matrix factorization techniques can also gain superior performance and scale to embeddings of large-scale graphs.

\noindent $\bullet$ {\em Scalable systems}.
Multi-CPU and multi-GPU platforms are widely-adopted scalable systems for distributed GNN training and graph embedding. Multi-CPU platforms enable distributed GNN training across multiple machines. Multi-GPU platforms employ CPU-GPU collaborative solutions, where GPUs conduct GNN training/ embedding, whereas CPUs handle computationally intensive tasks, including sampling, random walks, and workload partition. Modern hardware, e.g., FPGA, SSD, and ReRAM enable training larger graphs on a single machine, while providing accelerations, fault-awareness, and energy-efficiency.

\noindent $\bullet$ {\em Software-Hardware co-design}. PyTorch Geometric (PyG) and Deep Graph Library (DGL) are common software paradigms for GNN training. They support CPU and GPU computing, full-batch and mini-batch training, also provide APIs and user-defined functions to abstract computation and communication. Using them, more advanced software frameworks, e.g., AliGraph \cite{ZhuZYLZALZ19}, DistGNN \cite{MdMMMGHKAA21}, and DistDGL \cite{Zheng0WZSSGZK20} are developed which define user-friendly programming models (e.g., vertex-centric paradigm) and efficient data structures. They employ software-hardware co-design to reduce computation and communication costs via different parallelization schemes (e.g., pipeline parallelism), optimization strategies (e.g., synchronous vs. asynchronous communication, parameter server),
on-chip data reuse, etc.

\noindent $\bullet$ {\em Graph databases}. Popular graph databases (graph DBs), e.g., Neo4J, ArangoDB, Amazon Neptune, TigerGraph, and K\`{u}zu provide data science libraries and ML tools to support a number of graph embedding methods and GNN training \cite{Khan23}. Graph DB's disk-based storage systems can be used with PyG remote backend to train a GNN model on very large graphs that do not fit on the main memory of a single server \footnote{\scriptsize{{https://blog.kuzudb.com/post/kuzu-pyg-remote-backend/}}}. While graph DBs currently provide only basic graph ML functionalities such as node classification and regression, link prediction, it would be interesting to seamlessly integrate graph embeddings and GNN's capabilities into graph query processing and question answering (QA) (\S\ref{sec:kg}), also enabling vector indexes for efficient similarity search to facilitate graph RAG paradigm in LLMs (\S\ref{sec:graphrag}). These highlight the potential of graph DBs to be coupled with ML-based QA systems and LLMs \cite{abs-2410-03867}.

\noindent $\bullet$ {\em Improving graph data pipeline}. Finally, efficient graph embedding and GNN training are key to many downstream applications, e.g., graph data cleaning, entity resolution, and knowledge graph question answering, ensuring effective, efficient, and robust graph data pipelines.

\subsection{Graph-based Vector Data Indexes}
\label{sec:index}

\medskip
\medskip

The management of vector data intersects with graph data management, particularly in systems that support graph-based machine learning (GML).
A prime example is the use of graph-based vector indices, e.g., HNSW~\cite{DML11, MYD18HNSW, FXWC19} to organize high-dimensional embeddings for retrieval tasks.
These embeddings often originate from GML models like Graph Neural Networks (GNNs)~\cite{KLJZL20, BMHCRMKD22}, where node or graph-level representations are computed for downstream applications.
This synergy between graph-based indices and GML pipelines positions GDM systems, including Neo4j and TigerGraph, as comprehensive platforms for building GML workflows, integrating data storage, embedding generation, and similarity search functionalities.

Graph-based indices~\cite{PWL23Survey} diverge from traditional indexing methods, such as inverted indices~\cite{BL14II,jds10ii}, locality-sensitive hashing~\cite{AILRS15, TZZ23DB}, and tree-based indices~\cite{BCG05, KS18TB}, which typically partition vectors into buckets.
Instead, graph-based indices construct proximity graphs, where nodes represent data points and edges denote neighbor relationships.
These graph-based approaches present unparalleled effectiveness by leveraging semantic similarities through the principle that a neighbor's neighbor is likely to be a neighbor and iteratively expanding neighbors' neighbors through a best-first search~\cite{FanL19, wang2022crux}.
Recent works substantiate their scalability, positioning them for handling billion-scale datasets~\cite{WXYW21}.
Unlike traditional graph data structures used for representing networked information, these indices are optimized for the Approximate Nearest Neighbor Search (ANNS)~\cite{AIR18anns, LZSWLZL19, WXYW21}, a task foundational to many AI-driven applications.
This makes them particularly relevant to GDM systems that serve as infrastructure for hybrid tasks combining traditional graph analysis and ML-based embedding retrieval.
The implications of such methods extend beyond ANNS, permeating into the fabric of LLMs~\cite{KLJZL20} and unstructured data management~\cite{HHGDAH13, WWWLZLC20}, heralding a new era in the intersection of graph-based data management and real-world applications.

Graph-based vector indices have been subject to a range of optimizations aimed at improving both the index structure and search procedures, which can be categorized into four key areas:

The first major category, {\em graph index optimization}, focuses on diversifying neighbor connections to enhance graph navigability and capture semantic relationships between embeddings, such as refining the quality of the edge set~\cite{FXWC19}, leveraging more sophisticated distance functions~\cite{DML11}, adaptive neighbor selection~\cite{peng2023efficient}, and hierarchical layouts~\cite{MYD18HNSW}.
These ensure that similar embeddings are efficiently connected and easily discoverable during search.
The index graph quality directly impacts downstream GML tasks like node classification and link prediction, where effective and efficient similarity assessments are critical for model performance.

The second set of optimizations focuses on enhancing search strategies to reduce traversal overhead while maintaining high query accuracy. This is particularly important when scaling graph-based models to larger datasets, as the cost of inefficient traversal can quickly overwhelm the benefits of an optimized index.
{\em Routing optimizations} address this challenge by refining key aspects such as entry point acquisition~\cite{lu2021hvs, zhao2023towards}, routing strategy~\cite{gao2023high, yue2023routing, LXI24Pro, yue2023routing}, and termination conditions~\cite{li2020improving, zhang2023vbase}.
By combining these strategies, routing optimization ensures that even in large-scale GML datasets, searches remain fast and precise, minimizing the impact of increasing data size on performance.

Building on these search optimizations, the third category focuses on scaling solutions through {\em hardware-aware optimizations}, which adapt the index layout and search strategies to specific hardware capabilities~\cite{WXYWPKGXGX24}.
Graph-based methods have been implemented in {\em external memory} such as heterogeneous memory (HM)~\cite{ren2020hm} and solid-state disk (SSD)~\cite{jayaram2019diskann}, to scale the system beyond traditional memory limitations.
A recent work~\cite{jang2023cxl} has adapted graph-based indexes to the cutting-edge compute express link (CXL) architecture.
In addition, {\em acceleration hardware} such as GPUs and FPGAs are utilized to parallelize vector computation~\cite{zhao2020song, ootomo2024cagra, manohar2024parlayann} or data structure maintenance~\cite{yu2022gpu}, providing an order of magnitude increase in efficiency for both index construction and search,
These innovations exemplify how {\em software-hardware collaboration} enables scalable solutions for embedding-intensive GML tasks, addressing computational bottlenecks in GML workflows.

The fourth line of research integrates additional information into graph indices to further support more sophisticated retrieval scenarios, a critical need for complex graph ML workflows.
Techniques such as attribute-based filtering~\cite{wang2023efficient, gollapudi2023filtered, PKGZ24ACORN, ZQZLD4SERF} incorporate structured attributes directly into the index, enabling hybrid queries that combine structured and unstructured data.
For instance, in {\em multimodal search} scenarios, where each entity comprises multiple vectors, {\em multiple} graph indexes may be constructed and scanned to address a multi-vector query~\cite{wang2021milvus,zhang2023vbase}.
An innovative approach \cite{WKXCGHZ24} has fused multiple embeddings into a unified graph index with automatic weight learning, enabling efficient and accurate multimodal queries.
These methods have demonstrated applicability in ML-powered systems, such as LLM-based online query answering~\cite{WWKGXC24}, further bridging the gap between advanced data management and real-world applications.

\spara{Synergy.}
We illustrate how advancements in graph-based vector indices, a core GDM technique, significantly contribute to the scalability and efficacy of GML systems.

\noindent $\bullet$ {\em Efficient embedding management.}
Graph indices excel at managing high-dimensional embeddings generated by GML tasks, such as node classification and link prediction. By leveraging optimizations in graph structure and search procedures, including neighbor diversification, efficient routing, and hardware acceleration, these indices enable faster and more precise similarity searches essential for embedding-driven GML workflows.

\noindent $\bullet$ {\em Scalable multimodal integration.}
For multimodal GML tasks, where nodes or entities are represented by multiple embeddings, graph indices adapt to efficiently handle hybrid and multimodal queries. Techniques like fused graph indices allow simultaneous processing of multiple data modalities, directly benefiting use cases like multimodal knowledge retrieval and enhanced representation learning in large-scale systems.

\noindent $\bullet$ {\em Hardware acceleration for GML.}
The alignment of graph indices with emerging hardware architectures, such as GPUs, FPGAs, and CXL, drives substantial improvements in computation and memory efficiency. These optimizations enable graph ML systems to scale effectively, overcoming the limitations of traditional memory-based approaches for embedding-intensive workloads.

\noindent $\bullet$ {\em Enhanced machine learning pipelines.}
By integrating attribute filtering, handling incomplete data, and accommodating large-scale retrieval, graph indices bolster the robustness of GML pipelines. This ensures reliable and efficient data processing for tasks such as hybrid query answering, anomaly detection, and fair representation learning. The adaptability of graph indices to evolving GML requirements demonstrates their critical role in enabling complex, real-world applications.

\subsection{GNN Explainability}
\label{sec:explainability}

\medskip
\medskip

To safely and trustfully deploy deep neural models, it is critical to provide human-intelligible explanations to end users and domain experts: {\em Which aspects of the input data drive the decisions of the model?} Therefore, explainability methods for GNNs are becoming popular.

Deriving and comparing GNN explanations are difficult. {\bf (1)} There is no unique notion of explainability -- the requirements arise due to many factors, e.g., trust, causality, transferability, fair decision making,  model debugging, informativeness, etc. \cite{Lipton18,KimD21}. {\bf (2)} Analogously, several quantitative metrics
such as fidelity, sparsity, contrastivity, and stability are proposed to evaluate explanation quality. It may be required to modify these metrics to capture the complex dependency of structure and feature in the graph space. For instance,
perturbation-based metrics (e.g., fidelity) can drastically change the graph's structure, resulting in data outside the training distribution. Instead, a standard practice is to consider ``milder'' perturbations by removing associated features of important nodes and edges, while keeping the graph structure intact \cite{PKRMH18,yuan2022explainability}. {\bf (3)} Due to the emerging nature of graph data and downstream tasks, there has been less qualitative evaluation of GNN explainability (e.g., human grounded evaluation) \cite{VT20,abs-2206-13983}. Lack of real-world ground-truth explanations, complexity in graph data, and requirement of expert domain knowledge are key bottlenecks behind qualitative evaluation. {\bf (4)} The output of GNN explainability (e.g., nodes, edges, features, subgraphs) and their categories (e.g., factual vs. counterfactual, instance vs. model-level) are different. For a holistic evaluation, such factors must be considered \cite{KM23}. {\bf (5)} Other concerns include non-robust GNN models and training bias \cite{FMW21}.

Recently, many explainability methods for GNNs have been developed, which can be categorized across several dimensions \cite{yuan2022explainability,KJSAM23,KM23}.
{\em Self-explanatory} approaches incorporate explainability directly into GNN models, e.g., \cite{DW21,ZhangLWLL22}.
{\em Post-hoc} methods \cite{YingBYZL19,FKA21,LCXYZCZ20,YTHJ20,SCT21,VT20,YuanYWLJ21} create a separate model to provide explanations for an existing GNN.
In {\em global} explanation methods, users understand how the model works globally by inspecting the structures and parameters of a GNN model, or by generating graph patterns which maximize a certain prediction of the model \cite{YTHJ20}. In contrast, {\em local} methods examine an individual prediction of a model, figuring out why the model makes the decision on a specific test instance \cite{YingBYZL19,FKA21,SCT21,VT20,YuanYWLJ21}. {\em Forward} explainability methods are GNN model-agnostic by learning evidences about graphs or nodes passed through the GNN. They can be {\em perturbation-based}, that is, masking some node features and/or edge features and analyzing the changes when the modified graphs are passed through GNNs \cite{YingBYZL19}. They might also employ a simple, explainable {\em surrogate model} to approximate the predictions of a complex GNN \cite{VT20}. In contrast, {\em backward} interpretability methods are GNN model-specific and can be either {\em gradient-based} \cite{PKRMH18} -- backpropagating importance signals backward from the output neuron of the model to the individual nodes of the input graph, or {\em decomposition-based} \cite{SchnakeELNSMM22} -- distributing the prediction score in a backpropagation manner until the input layer. Thus, one identifies which nodes, edges, and features contribute the most to the specific output label in the GNN. Furthermore, GNN explainability methods can be classified as {\em factual} (i.e., finding a subgraph whose information is sufficient -- which, if retained, will result in the same prediction), {\em counterfactual} (i.e., finding a subgraph that is necessary -- which, if removed, will result in a different prediction), or both \cite{TGFGXLZ22}.

However, existing approaches in this field are limited to providing explanations for individual instances or specific class labels.
The main focus of these methods is on defining
explanations as crucial input features, often in the shape of numerical encoding.
These methods generally fall short in {\em providing targeted and configurable explanations for multiple class labels of interest}.
Additionally, existing methods may return large explanation structures
and hence are not easily comprehensible. These explanation structures often lack direct accessibility
and cannot be queried easily, posing a challenge for expert users who seek to
inspect the specific reasoning behind a GNN's decision based on domain knowledge.

A recent work, GVEX \cite{ChenQWKKG24} proposes a novel two-tier explanation
structure called {\em explanation views}. An explanation view (similar to {\em graph view}) comprises a collection of graph patterns along with a set of induced explanation subgraphs.
Given a database of multiple graphs and a specific class label assigned by a GNN-based classifier, lower-tier
subgraphs provide insights into the reasons behind the assignment of the label by the classifier. They
serve as both factual (that preserves the result of classification) and counterfactual explanations
(which flips the result if removed). On the other hand, the higher-tier patterns summarize the subgraphs
using common substructures for efficient search and exploration
of these subgraphs. Analogously, RoboGExp \cite{QWKW24} introduces a new class of explanation structures to provide robust, both counterfactual and factual explanations for graph neural networks. Given a GNN, a robust explanation refers to the fraction of
a graph that are counterfactual and factual explanation of the results of the GNN over the graph, but also remains so for any ``disturbance'' by flipping up to $k$ of its node pairs.
In particular, such explanation indicates ``invariant'' representative structures
for similar graphs that fall into the same group, i.e., be ``robust'' to small
changes of the graphs, and be both ``factual'' and
``counterfactual''. Both GVEX and RoboGExp also emphasize effective, efficient, and scalable explanation generation by providing theoretical approximation guarantees and developing parallel and streaming algorithms.

\spara{Synergy.} We depict how GDM assists in generating better GNN explanations, which is a GML task.

\noindent $\bullet$ {\em Useful explanations}. First, explanations should not only dissect the decision-making process of GNN models, but can also {\em zoom in/out} on how certain features, nodes, or subgraphs contribute to specific classifications, that is, explanations can be provided across multiple granularity of concept hierarchy depending on the needs of end users. Moreover, enhancing the {\em accessibility}, {\em configurability}, and {\em queryability} of explanations
is crucial. Graph view-based two-tier explanations in GVEX \cite{ChenQWKKG24} provide the first step in this direction, and a natural extension might be generating an explanation OLAP cube that can be drill up/down based on domain-specific requirements.
Second, explanations should be presented in a {\em user-friendly} manner, possibly through
visualizations or interactive tools that allow users to explore and interrogate GNNs' decisions. These tools could enable desirable capabilities, e.g., highlighting critical substructures, providing interactive interfaces, and allowing tunable parameters for domain experts to ``query'' the model about its decisions. It is paramount to think beyond ``explanation of GNN models'' and towards ``explanations for users'' to enable trust and effective deployment.

\noindent $\bullet$ {\em Efficient explanations}. Past research on explanation generation often does not emphasize on efficiency and scalability, e.g., requiring more than one day to generate an explanation over large-scale graphs \cite{ChenQWKKG24}. Real-time explanations are key to interactiveness, configurability, queryability, and in-depth exploration of GNNs' decision making process.
Parallel, streaming, and anytime algorithms, modern hardware, and software-hardware co-design have potentials to reduce explanation time.

\noindent $\bullet$ {\em Diversified explanations}.
As stated earlier, several quantitative metrics, e.g., fidelity, sparsity, contrastivity, and stability are designed to evaluate explanation quality; however, no single measure is the best. It is important to pursue explanations that optimize
multi-objective quality criteria, while also improving diversity. Concepts from databases, such as Pareto optimality and a skyline set
of explanatory subgraphs can be useful.

\noindent $\bullet$ {\em Better explanations to improve data pipeline}.
Finally, explanations can reveal unfairness in GNN's decision making process, detect anomalies and potential threats, help in model debugging, and  assist organizations in meeting compliance and regulations, thereby improving the robustness of graph data pipeline.

\section{Graph ML for Graph \\ Data Management}
\label{sec:gml4gdm}

\medskip
\medskip

We illustrate applications of graph machine learning and graph-based LLMs in knowledge graphs
query answering and other data science tasks.

\subsection{Knowledge Graphs Query Answering}
\label{sec:kg}

\medskip
\medskip

Query answering over datasets is an important data management task. We consider knowledge graph (KG) -- a graph-based data model to store facts -- denoted as $\langle$subject, predicate, object$\rangle$ triples, or a large-scale graph having nodes (subjects and objects) and edges (predicates) \cite{WeikumDRS21}. Querying KGs is critical for web search, semantic search, fact checking, and personal assistants. However, it is difficult due to their massive volume, heterogeneity, incompleteness, and schema flexibility. Additionally, a user's query (e.g., natural language query or query graph) may not match exactly w.r.t. entities, relations, and structure of the KG, requiring approximate matches to retrieve relevant answers \cite{Khan23}.

Machine learning assists in {\bf (1)} inferencing over KGs to identify missing relations during query answering, and also {\bf (2)} finding approximate matches for queries \cite{Abdallah023,HorchidanC23,abs-2303-14617}. {\bf (3)} Natural language queries (NLQs) are semantically parsed to structured queries (e.g., SPARQL queries over KGs) using neural approaches \cite{QuamarELO22}. {\bf (4)} More recent techniques employ sequential models for end-to-end answering of NLQs over KGs, e.g., KEQA \cite{HuangZLL19} for simple NLQs and EmbedKGQA \cite{SaxenaTT20} for multi-hop NLQs. {\bf (5)} KG embedding methods can be useful as well. Wang et al. \cite{WangKXJHF22,0001KWJY20} decompose multi-hop and complex queries into smaller subqueries, answer each subquery via single-hop reasoning with KG embedding, and then assemble the answers. In contrast,  Query2box \cite{RenHL20}  and follow-up works
train on multi-hop queries
– they embed multi-hop logic queries and their answers (i.e., entities
from a KG) in the same embedding space to reduce the query processing cost via inference.

Domain-specific knowledge graphs (KGs)~\cite{wang2022crux, li2020kg}
have been curated to host scientific,
factual knowledge rather than
generic Web or common knowledge,
such as KGs in material
science, healthcare, medicine, education,
cybersecurity, biology, and chemistry.
While knowledge curation has been extensively
studied, searching
domain data remains nontrivial.
Domain experts are still expected to
write complex declarative queries (such as
SPARQL), or data scripts to access KGs. There is a
gap between the need of accessing KGs with
(domain) languages
and optimized
query processing within
state-of-the-art KG data systems.
The rise of large language models (LLMs),
such as GPT provides promising capabilities in
generating natural language solutions in response to
users' prompts.
There are efforts on linking
LLMs to KG search and exploration~\cite{PanLWCWW24}, as well as LLM-based
knowledge graph
exploratory search~\cite{graphlingo}.
KG-enhanced, LLM-based QA is also studied: QAGNN \cite{YasunagaRBLL21} and GreaseLM \cite{ZBYRLML22} fine-tune a vanilla LM with a KG on downstream tasks, whereas DRAGON \cite{YasunagaBR0MLL22} and JAKET \cite{Yu0Y022} perform self-supervised pre-training from both text and KGs at scale.

\spara{Synergy.}
Query processing is the bread-and-butter for the data management community.
We highlight how GML and LLMs assist in KG querying and QA.

\noindent $\bullet$ {\em Natural language query processing}. Natural language interfaces to databases (NLIDB) is the holy grail for query interface to DBs -- automatically translating natural language questions (NLQs) to structured queries (e.g., SQL) that can be processed by a database management system. With the prevalence of graph data (e.g., domain-specific KGs) and the standardization of graph query languages (GQL), there is an emerging need to covert NLQs to graph queries, e.g., Cypher, SPARQL, Gremlin, GSQL, PGQL, etc. This is more challenging due to the complexity and expressivity of graph queries, coupled with the schema-flexibility and heterogeneity in graph data. GNNs and LLMs can assist in these tasks because of their understanding of contexts in conversational QA, background knowledge, and capability of dealing with natural language text. For instance, Neo4J recently developed NeoDash\footnote{\scriptsize{{https://neo4j.com/labs/neodash/2.4/user-guide/extensions/natural-language-queries/}}} which leverages LLMs to interpret user's input NLQs and generates Cypher queries based on the provided schema definition.

\noindent $\bullet$ {\em Approximate query processing}.
KGs are schema flexible, i.e., similar relationships between entity pairs can be represented in different ways. Therefore, one needs to construct various query patterns to retrieve all relevant answers from the underlying dataset, which is challenging. This necessitates approximate matches w.r.t. users' queries by understanding the query intent -- KG embedding and KG + query embedding approaches can support approximate matching via inference.

\noindent $\bullet$ {\em Query processing over incomplete data}.
KGs follow the open-world assumption, i.e., they are incomplete. To retrieve the complete set of answers for a given query, one must infer missing relations in KGs. In contrast, relational DBs generally follow the closed-world assumption with the presumption that all relevant knowledge is explicitly stored within the DB. Additionally, dealing with missing graph structure is more challenging than imputing missing feature values. ML-based link prediction and multi-hop inference techniques can be coupled with graph queries to resolve these problems.

\noindent $\bullet$ {\em Multimodal and multilingual data and queries}.
Entities and relations in a KG can have features with different data modalities, e.g., text, images, and multimedia data. Analogously, text data in node features and queries can be in different languages. Dense vector embedding of multimodal and multilingual data, obtained via deep models, provide a unique opportunity to query such heterogeneous data. Data management techniques can also contribute in querying vector data with high-dimensional indexes and join, leveraging modern hardware and geometric data processing.

\noindent $\bullet$ {\em Graph databases and query optimization}.
ML approaches, e.g., deep learning, reinforcement learning, and LLMs have shown promises in optimizing database queries and enhancing database administration functions such as query optimization, workload management, indexing, and storage layouts. Although there are recent developments in deep learning methods for graph pattern search and cardinality estimation \cite{ZYZLR21},
more work is needed in AI-facilitated graph databases and query optimization.
Graph ML algorithms could play a pivotal role in predicting access patterns, node importance, learning graph indexes based on query characteristics. By leveraging historical usage data and graph topology, these systems can autonomously adapt storage strategies and retrieval
mechanisms to match the evolving needs.

\subsection{Graph RAG-based LLMs in Data Science Applications}
\label{sec:graphrag}

\medskip
\medskip

LLMs which are a category of generative AI models and proficient at generating new text contents, offer a myriad of opportunities in data science by automating data analysis, manipulation, querying, and interpretation, as well as in code synthesis, digital assistants, finance, law, and education. Nevertheless, due to poor reasoning capacity, outdated or lack of domain knowledge, expensive re-training costs, and limited context lengths of LLMs, LLM-based data science pipelines often struggle with complex tasks -- they hallucinate, i.e., generate factually incorrect, or even harmful contents. To address these issues, KGs are used as background knowledge to enhance LLMs for downstream tasks. The questions are parsed to identify relevant subgraphs from KGs, then they are integrated and fused with LLMs based on knowledge integration, prompt augmentation, and retrieval augmented generation (RAG). This framework, known as graph RAG or KG-RAG \cite{XuCGWDWL24}, is increasingly becoming popular due to its ability
to capture the global context, compared to conventional RAG that retrieves knowledge from embeddings of textual chunks.

Recent works \cite{mavromatis2024gnnraggraphneuralretrieval,he2024g,WLRSZD24,WHBQRXS23} develop
KG-unified language models in a graph RAG style.
They can be broadly categorized into two groups according to the roles of KGs: {\bf (1)} KGs as background knowledge, and {\bf (2)} KGs as reasoning
guidelines. While the former only retrieves relevant subgraphs as contexts based on input questions, the later retrieves the most relevant paths adaptively to guide the LLM’s reasoning process \cite{SXTWLGSG24}. Graph RAG is further added within LLM-based agent systems to leverage structured knowledge for enhanced decision-making and problem-solving capabilities \cite{SunTLA24}.

\spara{Synergy}.
Besides GDM, effective text or vector processing may benefit graph RAG. For example, {\bf (1)} What is the proper data model to represent and feed the retrieved knowledge to the LLM? Options include prompt-based or embedding-based data model.
For the former, prompt engineering can be explored, such as serializing subgraphs to token sequences or $\langle$subject, predicate, object$\rangle$ triples, to best exploit LLMs'
ability of text (natural language) processing.
The latter can be better supported by vector databases (see \S~\ref{sec:index}).
{\bf (2)} How to design indexes, search algorithms, and systems for more complex and hybrid vector search, including graph traversal with vector retrieval? Those may require unifying graph DBs and vector DBs as external memory of LLMs. {\bf (3)} Graph query optimization, (explanatory) views, and provenance can help in making graph RAG
 better grounded by linking LLM response to
 factual knowledge at scale. {\bf (4)} Last but not least,  graph DBs may be used as ``semantic caches'' of LLMs by indexing previous question-answer pairs into a graph or vector space, enabling semantic
matching with new queries instead of more expensive LLM API calls.
These create new opportunities for GDM and broader data management
techniques to play
critical roles for graph RAG systems.

\section{Related Work}
\label{sec:related}

\medskip
\medskip

The closest to our work are surveys and tutorials on ML for data management
and data management for ML, emphasizing on relational
data and RDBMS \cite{ChaiWLNL23,
Kumar0017,Polyzotis0WZ17,HulsebosDSP23}.
However, graph data result in unique challenges to both data management and ML (\S\ref{sec:introduction}), justifying the importance of our survey.

Additionally, there are related surveys and tutorials on, e.g., graph representation learning \cite{CaiZC18,CuiWPZ19}, graph neural networks \cite{WuPCLZY21,ZhangCZ22,ma2021deep}, AI for data preparation \cite{Chai0FL23}, the role of graph data in graph ML \cite{abs-2309-10979}, distributed GNN training \cite{ShaoLGYLMZCC24}, explainable AI in data management \cite{PradhanLGS22}, ML explainability and robustness \cite{DattaFLLSW21}, LLM+KG \cite{PanLWCWW24}, and high-dimensional vector similarity search \cite{EchihabiPZ21}, etc.
However, none of them investigate the synergy of GDM and GML. To the best of our knowledge, ours is the first survey exploring the synergies between graph data management and graph ML over the end-to-end graph data pipeline. We hope that our survey will bridge the gap between these two popular domains -- GDM and GML, and would inspire others to work on the emerging graph data challenges at their intersection.

\section{Future Directions}
\label{sec:conclusions}

\medskip
\medskip

Future work can be in several directions.

\spara{Real-time Graph Learning and Inference}.
The integration of
spatiotemporal GNNs
and dynamic graphs would enable
the real-time decision that can rapidly
explore evolving nodes and links. This calls for
adaptive graph query processing and optimization,
online graph learning, and real-time inference at scale.
Graph analysis in finance, healthcare,
security, and
manufacturing will
benefit significantly from this capability.

\spara{Privacy-preserving Graph ML}.
As the usage of graph data expands, so does the concern for privacy and security. Future developments in the synergy between graph machine learning and data management could delve into advanced privacy-preserving techniques for graph data. This might involve the integration of federated learning approaches, differential privacy, or novel encryption methods tailored to the unique characteristics of graph structures. Ensuring the confidentiality of sensitive graph information, while still extracting valuable insights, poses an exciting challenge.

\spara{Robust Graph ML}.
GNNs can be sensitive under a set of link perturbations or adversarial attacks. ML communities have investigated several approaches on how to quantify and improve the robustness of graph learning, e.g.,  certifiable robustness. Data management techniques such as graph sparsification and cleaning can also be employed. In the past, data imputation and integration for graphs
have been extensively studied with the objective of data correctness and completeness, instead it would be interesting to clean graphs for optimizing the robustness of graph learning.

\spara{Unifying LLMs+KGs+Vector DBs}.
Knowledge bases such as KGs and data lakes support holistic integration for multimodal
data arriving from heterogeneous sources, including tabular, key-value pairs, text, images, and multimedia data. Vector embedding represents each predicate and entity from diverse sources as a low-dimensional vector,
such that the original structures and relations in the knowledge base are approximately preserved.
Querying these vectors are essential for a wide range of applications, e.g., question answering and
semantic search.
Finally, LLM pipelines are generally faster than traditional ML lifecycles --
thanks to simpler prompt-based interactions without any requirement of re-training, making it easy to build AI pipelines around LLMs. Thus, the unification of three modern technologies LLMs, KGs, and vector DBs seem indispensable.
There also remain many fundamental challenges, e.g.,  how to create a holistic embedding across multiple modalities and diverse data formats?
It remains a desirable yet nontrivial task to explain the results of LLMs and to incorporate domain knowledge -- KGs could assist in both objectives following graph RAG approaches. Analogously, adding human-in-the-loop and analyzing utility vs. privacy, bias, and fairness to derive quality solutions are important.

\section{Acknowledgment}
\label{sec:ack}

\medskip
\medskip

Khan acknowledges support from the Novo Nordisk Foundation grant NNF 22OC0072415. Ke is supported by Zhejiang Province's ``Lingyan'' R\&D Project under Grant No. 2024C01259 and Yongjiang Talent Introduction Programme (2022A-237-G). Wu is supported by NSF under CNS-1932574, CNS-2028748, and OAC-2104007.

\balance

\begin{small}
\bibliographystyle{abbrv}
\bibliography{ref}

\begin{thebibliography}{100}

\vspace{5mm}

\bibitem{Abdallah023}
H.~Abdallah and E.~Mansour.
\newblock {Towards A GML-enabled Knowledge Graph Platform}.
\newblock In {\em ICDE}, 2023.

\bibitem{AliBHVGSFTL22}
M.~Ali, M.~Berrendorf, C.~T. Hoyt, L.~Vermue, M.~Galkin, S.~Sharifzadeh,
  A.~Fischer, V.~Tresp, and J.~Lehmann.
\newblock {Bringing Light Into the Dark: A Large-Scale Evaluation of Knowledge
  Graph Embedding Models Under a Unified Framework}.
\newblock {\em {IEEE} Trans. Pattern Anal. Mach. Intell.}, 44(12):8825--8845,
  2022.

\bibitem{AILRS15}
A.~Andoni, P.~Indyk, T.~Laarhoven, I.~Razenshteyn, and L.~Schmidt.
\newblock {Practical and Optimal LSH for Angular Distance}.
\newblock In {\em NeurIPS}, 2015.

\bibitem{AIR18anns}
A.~Andoni, P.~Indyk, and I.~Razenshteyn.
\newblock {Approximate Nearest Neighbor Search in High Dimensions}.
\newblock In {\em International Congress of Mathematicians: Rio de Janeiro},
  2018.

\bibitem{ACPSSY23}
A.~Antelmi, G.~Cordasco, M.~Polato, V.~Scarano, C.~Spagnuolo, and D.~Yang.
\newblock {A Survey on Hypergraph Representation Learning}.
\newblock {\em ACM Computing Surveys}, 56(1):1--38, 2023.

\bibitem{BL14II}
A.~Babenko and V.~Lempitsky.
\newblock {The Inverted Multi-index}.
\newblock {\em IEEE Trans. on Pattern Anal. Mach. Intell.}, 37(6):1247--1260,
  2014.

\bibitem{BMVZ21}
C.~D.~T. Barros, M.~R.~F. Mendon\c{c}a, A.~B. Vieira, and A.~Ziviani.
\newblock {A Survey on Embedding Dynamic Graphs}.
\newblock {\em ACM Comput. Surv.}, 55(1), 2021.

\bibitem{BCG05}
M.~Bawa, T.~Condie, and P.~Ganesan.
\newblock {LSH Forest: Self-tuning Indexes for Similarity Search}.
\newblock In {\em WWW}, 2005.

\bibitem{BMHCRMKD22}
S.~Borgeaud, A.~Mensch, J.~Hoffmann, T.~Cai, E.~Rutherford, K.~Millican,
  G.~v.~d. Driessche, J.-B. Lespiau, B.~Damoc, A.~Clark, D.~d.~L. Casas,
  A.~Guy, J.~Menick, R.~Ring, T.~Hennigan, S.~Huang, L.~Maggiore, C.~Jones,
  A.~Cassirer, A.~Brock, M.~Paganini, G.~Irving, O.~Vinyals, S.~Simon~Osindero,
  K.~Simonyan, J.~W. Rae, E.~Elsen, and L.~Sifre.
\newblock {Improving Language Models by Retrieving from Trillions of Tokens}.
\newblock In {\em ICML}, 2022.

\bibitem{BrunaZSL13}
J.~Bruna, W.~Zaremba, A.~Szlam, and Y.~LeCun.
\newblock {Spectral Networks and Locally Connected Networks on Graphs}.
\newblock In {\em ICLR}, 2014.

\bibitem{CaiZC18}
H.~Cai, V.~W. Zheng, and K.~C. Chang.
\newblock {A Comprehensive Survey of Graph Embedding: Problems, Techniques, and
  Applications}.
\newblock {\em {IEEE} Trans. Knowl. Data Eng.}, 30(9):1616--1637, 2018.

\bibitem{CaiLWJ22}
L.~Cai, J.~Li, J.~Wang, and S.~Ji.
\newblock {Line Graph Neural Networks for Link Prediction}.
\newblock {\em {IEEE} Trans. Pattern Anal. Mach. Intell.}, 44(9):5103--5113,
  2022.

\bibitem{Chai0FL23}
C.~Chai, N.~Tang, J.~Fan, and Y.~Luo.
\newblock {Demystifying Artificial Intelligence for Data Preparation}.
\newblock In {\em SIGMOD}, 2023.

\bibitem{ChaiWLNL23}
C.~Chai, J.~Wang, Y.~Luo, Z.~Niu, and G.~Li.
\newblock {Data Management for Machine Learning: A Survey}.
\newblock {\em {IEEE} Trans. Knowl. Data Eng.}, 35(5):4646--4667, 2023.

\bibitem{ChenWDL00W20}
M.~Chen, Z.~Wei, B.~Ding, Y.~Li, Y.~Yuan, X.~Du, and J.~Wen.
\newblock {Scalable Graph Neural Networks via Bidirectional Propagation}.
\newblock In {\em NeurIPS}, 2020.

\bibitem{ChenQWKKG24}
T.~Chen, D.~Qiu, Y.~Wu, A.~Khan, X.~Ke, and Y.~Gao.
\newblock {View-based Explanations for Graph Neural Networks}.
\newblock {\em Proc. {ACM} Manag. Data}, 2(1):40:1--40:27, 2024.

\bibitem{CuiWPZ19}
P.~Cui, X.~Wang, J.~Pei, and W.~Zhu.
\newblock {A Survey on Network Embedding}.
\newblock {\em {IEEE} Trans. Knowl. Data Eng.}, 31(5):833--852, 2019.

\bibitem{DW21}
E.~Dai and S.~Wang.
\newblock {Towards Self-Explainable Graph Neural Network}.
\newblock In {\em CIKM}, 2021.

\bibitem{DattaFLLSW21}
A.~Datta, M.~Fredrikson, K.~Leino, K.~Lu, S.~Sen, and Z.~Wang.
\newblock {Machine Learning Explainability and Robustness: Connected at the
  Hip}.
\newblock In {\em KDD}, 2021.

\bibitem{DML11}
W.~Dong, C.~Moses, and K.~Li.
\newblock {Efficient k-nearest Neighbor Graph Construction for Generic
  Similarity Measures}.
\newblock In {\em WWW}, 2011.

\bibitem{DuvenaudMABHAA15}
D.~Duvenaud, D.~Maclaurin, J.~Aguilera{-}Iparraguirre,
  R.~G{\'{o}}mez{-}Bombarelli, T.~Hirzel, A.~Aspuru{-}Guzik, and R.~P. Adams.
\newblock {Convolutional Networks on Graphs for Learning Molecular
  Fingerprints}.
\newblock In {\em NeurIPS}, 2015.

\bibitem{EchihabiPZ21}
K.~Echihabi, T.~Palpanas, and K.~Zoumpatianos.
\newblock {New Trends in High-D Vector Similarity Search: AI-driven,
  Progressive, and Distributed}.
\newblock {\em PVLDB}, 14(12):3198--3201, 2021.

\bibitem{FMW21}
L.~Faber, A.~K. Moghaddam, and R.~Wattenhofer.
\newblock {When Comparing to Ground Truth is Wrong: On Evaluating GNN
  Explanation Methods}.
\newblock In {\em KDD}, 2021.

\bibitem{FanFTD15}
W.~Fan, Z.~Fan, C.~Tian, and X.~L. Dong.
\newblock {Keys for Graphs}.
\newblock {\em PVLDB}, 8(12):1590--1601, 2015.

\bibitem{FanJLLTZ20}
W.~Fan, R.~Jin, M.~Liu, P.~Lu, C.~Tian, and J.~Zhou.
\newblock {Capturing Associations in Graphs}.
\newblock {\em PVLDB}, 13(11):1863--1876, 2020.

\bibitem{FanL19}
W.~Fan and P.~Lu.
\newblock {Dependencies for Graphs}.
\newblock {\em TODS}, 44(2):5:1--5:40, 2019.

\bibitem{FangKLWFLYC23}
P.~Fang, A.~Khan, S.~Luo, F.~Wang, D.~Feng, Z.~Li, W.~Yin, and Y.~Cao.
\newblock {Distributed Graph Embedding with Information-Oriented Random Walks}.
\newblock {\em PVLDB}, 16(7):1643--1656, 2023.

\bibitem{FXWC19}
C.~Fu, C.~Xiang, C.~Wang, and D.~Cai.
\newblock {Fast Approximate Nearest Neighbor Search with the Navigating
  Spreading-out Graph}.
\newblock {\em PVLDB}, 12:461--–474, 2019.

\bibitem{FKA21}
T.~Funke, M.~Khosla, M.~Rathee, and A.~Anand.
\newblock {Zorro: Valid, Sparse, and Stable Explanations in Graph Neural
  Networks}.
\newblock {\em {IEEE} Trans. Knowl. Data Eng.}, 35(8):8687--8698, 2023.

\bibitem{gao2023high}
J.~Gao and C.~Long.
\newblock {High-Dimensional Approximate Nearest Neighbor Search: with Reliable
  and Efficient Distance Comparison Operations}.
\newblock {\em PACMMOD}, 1(2):1--27, 2023.

\bibitem{gollapudi2023filtered}
S.~Gollapudi, N.~Karia, V.~Sivashankar, R.~Krishnaswamy, N.~Begwani, S.~Raz,
  Y.~Lin, Y.~Zhang, N.~Mahapatro, P.~Srinivasan, et~al.
\newblock {Filtered-DiskANN: Graph Algorithms for Approximate Nearest Neighbor
  Search with Filters}.
\newblock In {\em WWW}, 2023.

\bibitem{GroverL16}
A.~Grover and J.~Leskovec.
\newblock {Node2vec: Scalable Feature Learning for Networks}.
\newblock In {\em KDD}, 2016.

\bibitem{gdet}
S.~Guan, H.~Ma, P.~Lin, and Y.~Wu.
\newblock {GEDet: Adversarially Learned Few-shot Detection of Erroneous Nodes
  in Graphs}.
\newblock In {\em IEEE BigData}, 2020.

\bibitem{guan2023gale}
S.~Guan, H.~Ma, M.~Wang, and Y.~Wu.
\newblock {GALE: Active Adversarial Learning for Erroneous Node Detection in
  Graphs}.
\newblock In {\em ICDE}, 2023.

\bibitem{GuptaGLSWZ13}
P.~Gupta, A.~Goel, J.~Lin, A.~Sharma, D.~Wang, and R.~Zadeh.
\newblock {WTF: The Who to Follow Service at Twitter}.
\newblock In {\em WWW}, 2013.

\bibitem{HamiltonYL17}
W.~L. Hamilton, Z.~Ying, and J.~Leskovec.
\newblock {Inductive Representation Learning on Large Graphs}.
\newblock In {\em NeurIPS}, 2017.

\bibitem{HanJLH22}
X.~Han, Z.~Jiang, N.~Liu, and X.~Hu.
\newblock {G-Mixup: Graph Data Augmentation for Graph Classification}.
\newblock In {\em ICML}, 2022.

\bibitem{he2024g}
X.~He, Y.~Tian, Y.~Sun, N.~V. Chawla, T.~Laurent, Y.~LeCun, X.~Bresson, and
  B.~Hooi.
\newblock {G-Retriever: Retrieval-Augmented Generation for Textual Graph
  Understanding and Question Answering}.
\newblock {\em arXiv preprint arXiv:2402.07630}, 2024.

\bibitem{HorchidanC23}
S.~Horchidan and P.~Carbone.
\newblock {ORB: Empowering Graph Queries through Inference}.
\newblock In {\em Joint Proceedings of the {ESWC} 2023 Workshops and
  Tutorials}, 2023.

\bibitem{HuCHFL17}
J.~Hu, R.~Cheng, Z.~Huang, Y.~Fang, and S.~Luo.
\newblock {On Embedding Uncertain Graphs}.
\newblock In {\em CIKM}, 2017.

\bibitem{HHGDAH13}
P.-S. Huang, X.~He, J.~Gao, L.~Deng, A.~Acero, and L.~Heck.
\newblock {Learning Deep Structured Semantic Models for Web Search Using
  Clickthrough Data}.
\newblock In {\em CIKM}, 2013.

\bibitem{HuangZLL19}
X.~Huang, J.~Zhang, D.~Li, and P.~Li.
\newblock {Knowledge Graph Embedding Based Question Answering}.
\newblock In {\em WSDM}, 2019.

\bibitem{HulsebosDSP23}
M.~Hulsebos, X.~Deng, H.~Sun, and P.~Papotti.
\newblock {Models and Practice of Neural Table Representations}.
\newblock In {\em SIGMOD}, 2023.

\bibitem{jang2023cxl}
J.~Jang, H.~Choi, H.~Bae, S.~Lee, M.~Kwon, and M.~Jung.
\newblock {CXL-ANNS: Software-Hardware Collaborative Memory Disaggregation and
  Computation for Billion-Scale Approximate Nearest Neighbor Search}.
\newblock In {\em USENIX ATC}, 2023.

\bibitem{jayaram2019diskann}
S.~Jayaram~Subramanya, F.~Devvrit, H.~V. Simhadri, R.~Krishnawamy, and
  R.~Kadekodi.
\newblock {Diskann: Fast Accurate Billion-point Nearest Neighbor Search on a
  Single Node}.
\newblock {\em NeurIPS}, 32, 2019.

\bibitem{jds10ii}
H.~Jegou, M.~Douze, and C.~Schmid.
\newblock {Product Quantization for Nearest Neighbor Search}.
\newblock {\em {IEEE} Trans. Pattern Anal. Mach. Intell.}, 33(1):117--128,
  2010.

\bibitem{jin2023large}
B.~Jin, G.~Liu, C.~Han, M.~Jiang, H.~Ji, and J.~Han.
\newblock {Large Language Models on Graphs: A Comprehensive Survey}.
\newblock {\em {IEEE} Trans. Knowl. Data Eng.}, 36(12):8622--8642, 2024.

\bibitem{JinZZLTS22}
W.~Jin, L.~Zhao, S.~Zhang, Y.~Liu, J.~Tang, and N.~Shah.
\newblock {Graph Condensation for Graph Neural Networks}.
\newblock In {\em ICLR}, 2022.

\bibitem{JBM24}
P.~Juillard, A.~Bonifati, and A.~Maur.
\newblock {Interactive Graph Repairs for Neighborhood Constraints}.
\newblock In {\em EDBT}, 2024.

\bibitem{KJSAM23}
J.~Kakkad, J.~Jannu, K.~Sharma, C.~C. Aggarwal, and S.~Medya.
\newblock {A Survey on Explainability of Graph Neural Networks}.
\newblock {\em {IEEE} Data Eng. Bull.}, 46(2):35--63, 2023.

\bibitem{KS18TB}
O.~Keivani and K.~Sinha.
\newblock {Improved Nearest Neighbor Search Using Auxiliary Information and
  Priority Functions}.
\newblock In {\em ICML}, 2018.

\bibitem{Khan23}
A.~Khan.
\newblock {Knowledge Graphs Querying}.
\newblock {\em {SIGMOD} Rec.}, 52(2):18--29, 2023.

\bibitem{KM23}
A.~Khan and E.~B. Mobaraki.
\newblock {Interpretability Methods for Graph Neural Networks}.
\newblock In {\em DSAA}, 2023.

\bibitem{KLJZL20}
U.~Khandelwal, O.~Levy, D.~Jurafsky, L.~Zettlemoyer, and M.~Lewis.
\newblock {Generalization through Memorization: Nearest Neighbor Language
  Models}.
\newblock In {\em ICLR}, 2020.

\bibitem{KimD21}
B.~Kim and F.~Doshi{-}Velez.
\newblock {Machine Learning Techniques for Accountability}.
\newblock {\em {AI} Mag.}, 42(1):47--52, 2021.

\bibitem{KW16}
T.~N. Kipf and M.~Welling.
\newblock {Variational Graph Auto-Encoders}.
\newblock In {\em NeurIPS Workshop on Bayesian Deep Learning}, 2016.

\bibitem{KipfW17}
T.~N. Kipf and M.~Welling.
\newblock {Semi-Supervised Classification with Graph Convolutional Networks}.
\newblock In {\em ICLR}, 2017.

\bibitem{KlicperaBG19}
J.~Klicpera, A.~Bojchevski, and S.~G{\"{u}}nnemann.
\newblock {Predict then Propagate: Graph Neural Networks meet Personalized
  PageRank}.
\newblock In {\em ICLR}, 2019.

\bibitem{Kumar0017}
A.~Kumar, M.~Boehm, and J.~Yang.
\newblock {Data Management in Machine Learning: Challenges, Techniques, and
  Systems}.
\newblock In {\em SIGMOD}, 2017.

\bibitem{graphlingo}
D.~Le, K.~Zhao, M.~Wang, and Y.~Wu.
\newblock {GraphLingo: Domain Knowledge Exploration by Synchronizing Knowledge
  Graphs and Large Language Models}.
\newblock In {\em ICDE}, 2024.

\bibitem{LererWSLWBP19}
A.~Lerer, L.~Wu, J.~Shen, T.~Lacroix, L.~Wehrstedt, A.~Bose, and
  A.~Peysakhovich.
\newblock {Pytorch-BigGraph: A Large Scale Graph Embedding System}.
\newblock In {\em MLSys}, 2019.

\bibitem{Li0SZAW20}
B.~Li, W.~Wang, Y.~Sun, L.~Zhang, M.~A. Ali, and Y.~Wang.
\newblock {GraphER: Token-Centric Entity Resolution with Graph Convolutional
  Neural Networks}.
\newblock In {\em AAAI}, 2020.

\bibitem{li2020improving}
C.~Li, M.~Zhang, D.~G. Andersen, and Y.~He.
\newblock {Improving Approximate Nearest Neighbor Search through Learned
  Adaptive Early Termination}.
\newblock In {\em SIGMOD}, 2020.

\bibitem{LZSWLZL19}
W.~Li, Y.~Zhang, Y.~Sun, W.~Wang, M.~Li, W.~Zhang, and X.~Lin.
\newblock {Approximate Nearest Neighbor Search on High Dimensional
  Data—Experiments, Analyses, and Improvement}.
\newblock {\em {IEEE} Trans. Knowl. Data Eng.}, 32(8):1475--1488, 2019.

\bibitem{abs-2311-12399}
Y.~Li, Z.~Li, P.~Wang, J.~Li, X.~Sun, H.~Cheng, and J.~X. Yu.
\newblock {A Survey of Graph Meets Large Language Model: Progress and Future
  Directions}.
\newblock In {\em IJCAI}, 2024.

\bibitem{li2020kg}
Y.~Li, V.~Zakhozhyi, D.~Zhu, and L.~J. Salazar.
\newblock {Domain Specific Knowledge Graphs as a Service to the Public}.
\newblock In {\em KDD}, 2020.

\bibitem{LinYYFPCX23}
H.~Lin, M.~Yan, X.~Ye, D.~Fan, S.~Pan, W.~Chen, and Y.~Xie.
\newblock {A Comprehensive Survey on Distributed Training of Graph Neural
  Networks}.
\newblock {\em Proc. {IEEE}}, 111(12):1572--1606, 2023.

\bibitem{linrepairing}
P.~Lin, Q.~Song, Y.~Wu, and J.~Pi.
\newblock {Repairing Entities using Star Constraints in Multirelational
  Graphs}.
\newblock In {\em ICDE}, 2020.

\bibitem{LinPCX17}
X.~Lin, Y.~Peng, B.~Choi, and J.~Xu.
\newblock {Human-Powered Data Cleaning for Probabilistic Reachability Queries
  on Uncertain Graphs}.
\newblock {\em {IEEE} Trans. Knowl. Data Eng.}, 29(7):1452--1465, 2017.

\bibitem{Lipton18}
Z.~C. Lipton.
\newblock {The Mythos of Model Interpretability}.
\newblock {\em Commun. {ACM}}, 61(10), 2018.

\bibitem{LLJLZDXL22}
C.~Liu, H.~Liu, H.~Jin, X.~Liao, Y.~Zhang, Z.~Duan, J.~Xu, and H.~Li.
\newblock {ReGNN: A ReRAM-Based Heterogeneous Architecture for General Graph
  Neural Networks}.
\newblock In {\em Proceedings of the 59th ACM/IEEE Design Automation
  Conference}, 2022.

\bibitem{abs-2310-11829}
J.~Liu, C.~Yang, Z.~Lu, J.~Chen, Y.~Li, M.~Zhang, T.~Bai, Y.~Fang, L.~Sun,
  P.~S. Yu, and C.~Shi.
\newblock {Towards Graph Foundation Models: A Survey and Beyond}.
\newblock {\em CoRR}, abs/2310.11829, 2023.

\bibitem{lu2021hvs}
K.~Lu, M.~Kudo, C.~Xiao, and Y.~Ishikawa.
\newblock {HVS: Hierarchical Graph Structure Based on Voronoi Diagrams for
  Solving Approximate Nearest Neighbor Search}.
\newblock {\em PVLDB}, 15(2):246--258, 2021.

\bibitem{LXI24Pro}
K.~Lu, C.~Xiao, and Y.~Ishikawa.
\newblock {Probabilistic Routing for Graph-Based Approximate Nearest Neighbor
  Search}.
\newblock In {\em ICML}, 2024.

\bibitem{LCXYZCZ20}
D.~Luo, W.~Cheng, D.~Xu, W.~Yu, B.~Zong, H.~Chen, and X.~Zhang.
\newblock {Parameterized Explainer for Graph Neural Network}.
\newblock In {\em NeurIPS}, 2020.

\bibitem{ma2021deep}
Y.~Ma and J.~Tang.
\newblock {\em {Deep Learning on Graphs}}.
\newblock Cambridge University Press, 2021.

\bibitem{MYD18HNSW}
Y.~A. Malkov and D.~A. Yashunin.
\newblock {Efficient and Robust Approximate Nearest Neighbor Search Using
  Hierarchical Navigable Small World Graphs}.
\newblock {\em {IEEE} Trans. Pattern Anal. Mach. Intell.}, 42(4):824--836,
  2018.

\bibitem{manohar2024parlayann}
M.~D. Manohar, Z.~Shen, G.~Blelloch, L.~Dhulipala, Y.~Gu, H.~V. Simhadri, and
  Y.~Sun.
\newblock {ParlayANN: Scalable and Deterministic Parallel Graph-Based
  Approximate Nearest Neighbor Search Algorithms}.
\newblock In {\em PPoPP24}, 2024.

\bibitem{mavromatis2024gnnraggraphneuralretrieval}
C.~Mavromatis and G.~Karypis.
\newblock {GNN-RAG: Graph Neural Retrieval for Large Language Model Reasoning}.
\newblock {\em CoRR}, 2024.

\bibitem{abs-2202-08455}
E.~Min, R.~Chen, Y.~Bian, T.~Xu, K.~Zhao, W.~Huang, P.~Zhao, J.~Huang,
  S.~Ananiadou, and Y.~Rong.
\newblock {Transformer for Graphs: An Overview from Architecture Perspective}.
\newblock {\em CoRR}, abs/2202.08455, 2022.

\bibitem{MohoneyWXRV21}
J.~Mohoney, R.~Waleffe, H.~Xu, T.~Rekatsinas, and S.~Venkataraman.
\newblock {Marius: Learning Massive Graph Embeddings on a Single Machine}.
\newblock In {\em OSDI}, 2021.

\bibitem{ootomo2024cagra}
H.~Ootomo, A.~Naruse, C.~Nolet, R.~Wang, T.~Feher, and Y.~Wang.
\newblock {CAGRA: Highly Parallel Graph Construction and Approximate Nearest
  Neighbor Search for GPUs}.
\newblock In {\em ICDE}, 2024.

\bibitem{PWL23Survey}
J.~J. Pan, J.~Wang, and G.~Li.
\newblock {Survey of Vector Database Management Systems}.
\newblock {\em {VLDB} J.}, 33(5):1591--1615, 2024.

\bibitem{PanLWCWW24}
S.~Pan, L.~Luo, Y.~Wang, C.~Chen, J.~Wang, and X.~Wu.
\newblock {Unifying Large Language Models and Knowledge Graphs: A Roadmap}.
\newblock {\em {IEEE} Trans. Knowl. Data Eng.}, 36(7):3580--3599, 2024.

\bibitem{abs-2410-03867}
R.~D. Pasquale and S.~Represa.
\newblock {Empowering Domain-Specific Language Models with Graph-Oriented
  Databases: A Paradigm Shift in Performance and Model Maintenance}.
\newblock {\em CoRR}, abs/2410.03867, 2024.

\bibitem{PKGZ24ACORN}
L.~Patel, P.~Kraft, C.~Guestrin, and M.~Zaharia.
\newblock {ACORN: Performant and Predicate-Agnostic Search Over Vector
  Embeddings and Structured Data}.
\newblock In {\em SIGMOD}, 2024.

\bibitem{peng2023efficient}
Y.~Peng, B.~Choi, T.~N. Chan, J.~Yang, and J.~Xu.
\newblock {Efficient Approximate Nearest Neighbor Search in Multi-dimensional
  Databases}.
\newblock {\em PACMMOD}, 1(1):1--27, 2023.

\bibitem{Polyzotis0WZ17}
N.~Polyzotis, S.~Roy, S.~E. Whang, and M.~Zinkevich.
\newblock {Data Management Challenges in Production Machine Learning}.
\newblock In {\em SIGMOD}, 2017.

\bibitem{PKRMH18}
P.~E. Pope, S.~Kolouri, M.~Rostami, C.~E. Martin, and H.~Hoffmann.
\newblock {Explainability Methods for Graph Convolutional Neural Networks}.
\newblock In {\em CVPR}, 2019.

\bibitem{PradhanLGS22}
R.~Pradhan, A.~Lahiri, S.~Galhotra, and B.~Salimi.
\newblock {Explainable AI: Foundations, Applications, Opportunities for Data
  Management Research}.
\newblock In {\em ICDE}, 2022.

\bibitem{QWKW24}
D.~Qiu, M.~Wang, A.~Khan, and Y.~Wu.
\newblock {Generating Robust Counterfactual Witnesses for Graph Neural
  Networks}.
\newblock In {\em ICDE}, 2024.

\bibitem{QiuDMLWT18}
J.~Qiu, Y.~Dong, H.~Ma, J.~Li, K.~Wang, and J.~Tang.
\newblock {Network Embedding as Matrix Factorization: Unifying DeepWalk, LINE,
  PTE, and Node2vec}.
\newblock In {\em WSDM}, 2018.

\bibitem{QuamarELO22}
A.~Quamar, V.~Efthymiou, C.~Lei, and F.~{\"{O}}zcan.
\newblock {Natural Language Interfaces to Data}.
\newblock {\em Found. Trends Databases}, 11(4):319--414, 2022.

\bibitem{rabbani2023shactor}
K.~Rabbani, M.~Lissandrini, and K.~Hose.
\newblock Shactor: improving the quality of large-scale knowledge graphs with
  validating shapes.
\newblock In {\em Companion of the International Conference on Management of
  Data (SIGMOD)}, pages 151--154, 2023.

\bibitem{abs-2206-13983}
M.~Rathee, T.~Funke, A.~Anand, and M.~Khosla.
\newblock {Bagel: A Benchmark for Assessing Graph Neural Network Explanations}.
\newblock {\em CoRR}, abs/2206.13983, 2022.

\bibitem{abs-2303-14617}
H.~Ren, M.~Galkin, M.~Cochez, Z.~Zhu, and J.~Leskovec.
\newblock {Neural Graph Reasoning: Complex Logical Query Answering Meets Graph
  Databases}.
\newblock {\em CoRR}, abs/2303.14617, 2023.

\bibitem{RenHL20}
H.~Ren, W.~Hu, and J.~Leskovec.
\newblock {Query2box: Reasoning over Knowledge Graphs in Vector Space Using Box
  Embeddings}.
\newblock In {\em ICLR}, 2020.

\bibitem{ren2020hm}
J.~Ren, M.~Zhang, and D.~Li.
\newblock {HM-ANN:} efficient billion-point nearest neighbor search on
  heterogeneous memory.
\newblock In {\em NeurIPS}, 2020.

\bibitem{SaxenaTT20}
A.~Saxena, A.~Tripathi, and P.~P. Talukdar.
\newblock {Improving Multi-hop Question Answering over Knowledge Graphs using
  Knowledge Base Embeddings}.
\newblock In {\em ACL}, 2020.

\bibitem{SCT21}
M.~S. Schlichtkrull, N.~D. Cao, and I.~Titov.
\newblock {Interpreting Graph Neural Networks for NLP with Differentiable Edge
  Masking}.
\newblock In {\em ICLR}, 2021.

\bibitem{SchlichtkrullKB18}
M.~S. Schlichtkrull, T.~N. Kipf, P.~Bloem, R.~van~den Berg, I.~Titov, and
  M.~Welling.
\newblock {Modeling Relational Data with Graph Convolutional Networks}.
\newblock In {\em ESWC}, 2018.

\bibitem{SchnakeELNSMM22}
T.~Schnake, O.~Eberle, J.~Lederer, S.~Nakajima, K.~T. Sch{\"{u}}tt,
  K.~M{\"{u}}ller, and G.~Montavon.
\newblock {Higher-Order Explanations of Graph Neural Networks via Relevant
  Walks}.
\newblock {\em {IEEE} Trans. Pattern Anal. Mach. Intell.}, 44(11):7581--7596,
  2022.

\bibitem{ShaoLGYLMZCC24}
Y.~Shao, H.~Li, X.~Gu, H.~Yin, Y.~Li, X.~Miao, W.~Zhang, B.~Cui, and L.~Chen.
\newblock {Distributed Graph Neural Network Training: A Survey}.
\newblock {\em {ACM} Comput. Surv.}, 56(8):191:1--191:39, 2024.

\bibitem{SiGXYDP22}
J.~Si, X.~Gan, T.~Xiao, B.~Yang, D.~Dong, and Z.~Pang.
\newblock {STEGNN: Spatial-Temporal Embedding Graph Neural Networks for Road
  Network Forecasting}.
\newblock In {\em ICPADS}, 2022.

\bibitem{SXTWLGSG24}
J.~Sun, C.~Xu, L.~Tang, S.~Wang, C.~Lin, Y.~Gong, H.~Shum, and J.~Guo.
\newblock {Think-on-Graph: Deep and Responsible Reasoning of Large Language
  Model with Knowledge Graph}.
\newblock In {\em ICLR}, 2024.

\bibitem{Sun00CXWY18}
L.~Sun, L.~He, Z.~Huang, B.~Cao, C.~Xia, X.~Wei, and P.~S. Yu.
\newblock {Joint Embedding of Meta-Path and Meta-Graph for Heterogeneous
  Information Networks}.
\newblock In {\em ICBK}, 2018.

\bibitem{SunTLA24}
L.~Sun, Z.~Tao, Y.~Li, and H.~Arakawa.
\newblock {ODA: Observation-Driven Agent for integrating LLMs and Knowledge
  Graphs}.
\newblock In {\em ACL}, 2024.

\bibitem{SureshLHN21}
S.~Suresh, P.~Li, C.~Hao, and J.~Neville.
\newblock {Adversarial Graph Augmentation to Improve Graph Contrastive
  Learning}.
\newblock In {\em NeurIPS}, 2021.

\bibitem{TGFGXLZ22}
J.~Tan, S.~Geng, Z.~Fu, Y.~Ge, S.~Xu, Y.~Li, and Y.~Zhang.
\newblock {Learning and Evaluating Graph Neural Network Explanations Based on
  Counterfactual and Factual Reasoning}.
\newblock In {\em Web Conference}, 2022.

\bibitem{TWCWYZL24}
D.~Tang, J.~Wang, R.~Chen, L.~Wang, W.~Yu, J.~Zhou, and K.~Li.
\newblock {XGNN: Boosting Multi-GPU GNN Training via Global GNN MemoryStore}.
\newblock {\em PVLDB}, 17(5):1105--1118, 2024.

\bibitem{abs-2310-13023}
J.~Tang, Y.~Yang, W.~Wei, L.~Shi, L.~Su, S.~Cheng, D.~Yin, and C.~Huang.
\newblock {GraphGPT: Graph Instruction Tuning for Large Language Models}.
\newblock In {\em SIGIR}, pages 491--500, 2024.

\bibitem{Tian22}
Y.~Tian.
\newblock The world of graph databases from an industry perspective.
\newblock {\em {SIGMOD} Rec.}, 51(4):60--67, 2022.

\bibitem{TZZ23DB}
Y.~Tian, X.~Zhao, and X.~Zhou.
\newblock {DB-LSH 2.0: Locality-Sensitive Hashing With Query-Based Dynamic
  Bucketing}.
\newblock {\em {IEEE} Trans. Knowl. Data Eng.}, 2023.

\bibitem{MdMMMGHKAA21}
M.~Vasimuddin, S.~Misra, G.~Ma, R.~Mohanty, E.~Georganas, A.~Heinecke, D.~D.
  Kalamkar, N.~K. Ahmed, and S.~Avancha.
\newblock Distgnn: Scalable distributed training for large-scale graph neural
  networks.
\newblock In {\em SC}, 2021.

\bibitem{VelickovicCCRLB18}
P.~Velickovic, G.~Cucurull, A.~Casanova, A.~Romero, P.~Li{\`{o}}, and
  Y.~Bengio.
\newblock {Graph Attention Networks}.
\newblock In {\em ICLR}, 2018.

\bibitem{VelickovicYPHB20}
P.~Velickovic, R.~Ying, M.~Padovano, R.~Hadsell, and C.~Blundell.
\newblock {Neural Execution of Graph Algorithms}.
\newblock In {\em ICLR}, 2020.

\bibitem{VT20}
M.~N. Vu and M.~T. Thai.
\newblock {PGM-Explainer: Probabilistic Graphical Model Explanations for Graph
  Neural Networks}.
\newblock In {\em NeurIPS}, 2020.

\bibitem{WangWWZZZXG18}
H.~Wang, J.~Wang, J.~Wang, M.~Zhao, W.~Zhang, F.~Zhang, X.~Xie, and M.~Guo.
\newblock {GraphGAN: Graph Representation Learning With Generative Adversarial
  Nets}.
\newblock In {\em AAAI}, 2018.

\bibitem{WangHZZZL18}
J.~Wang, P.~Huang, H.~Zhao, Z.~Zhang, B.~Zhao, and D.~L. Lee.
\newblock {Billion-scale Commodity Embedding for E-commerce Recommendation in
  Alibaba}.
\newblock In {\em KDD}, 2018.

\bibitem{wang2021milvus}
J.~Wang, X.~Yi, R.~Guo, H.~Jin, P.~Xu, S.~Li, X.~Wang, X.~Guo, C.~Li, X.~Xu,
  et~al.
\newblock {Milvus: A Purpose-built Vector Data Management System}.
\newblock In {\em SIGMOD}, 2021.

\bibitem{WKXCGHZ24}
M.~Wang, X.~Ke, X.~Xu, L.~Chen, Y.~Gao, et~al.
\newblock {MUST: An Effective and Scalable Framework for Multimodal Search of
  Target Modality}.
\newblock In {\em ICDE}, 2024.

\bibitem{wang2023efficient}
M.~Wang, L.~Lv, X.~Xu, Y.~Wang, Q.~Yue, and J.~Ni.
\newblock {An Efficient and Robust Framework for Approximate Nearest Neighbor
  Search with Attribute Constraint}.
\newblock In {\em NeurIPS}, 2023.

\bibitem{wang2022crux}
M.~Wang, H.~Ma, A.~Daundkar, S.~Guan, Y.~Bian, A.~Sehirlioglu, and Y.~Wu.
\newblock {CRUX: Crowdsourced Materials Science Resource and Workflow
  Exploration}.
\newblock In {\em CIKM}, 2022.

\bibitem{WWKGXC24}
M.~Wang, H.~Wu, X.~Ke, Y.~Gao, X.~Xu, and L.~Chen.
\newblock {An Interactive Multi-modal Query Answering System with
  Retrieval-Augmented Large Language Models}.
\newblock {\em PVLDB}, 17(12):1643--1656, 2024.

\bibitem{WXYWPKGXGX24}
M.~Wang, W.~Xu, X.~Yi, S.~Wu, Z.~Peng, X.~Ke, Y.~Gao, X.~Xu, R.~Guo, and
  C.~Xie.
\newblock {Starling: An I/O-Efficient Disk-Resident Graph Index Framework for
  High-Dimensional Vector Similarity Search on Data Segment}.
\newblock In {\em SIGMOD}, 2024.

\bibitem{WXYW21}
M.~Wang, X.~Xu, Q.~Yue, and Y.~Wang.
\newblock {A Comprehensive Survey and Experimental Comparison of Graph-based
  Approximate Nearest Neighbor Search}.
\newblock {\em PVLDB}, 14(11):1964–--1978, 2021.

\bibitem{0001KWJY20}
Y.~Wang, A.~Khan, T.~Wu, J.~Jin, and H.~Yan.
\newblock {Semantic Guided and Response Times Bounded Top-k Similarity Search
  over Knowledge Graphs}.
\newblock In {\em ICDE}, 2020.

\bibitem{WangKXJHF22}
Y.~Wang, A.~Khan, X.~Xu, J.~Jin, Q.~Hong, and T.~Fu.
\newblock {Aggregate Queries on Knowledge Graphs: Fast Approximation with
  Semantic-aware Sampling}.
\newblock In {\em ICDE}, 2022.

\bibitem{WLRSZD24}
Y.~Wang, N.~Lipka, R.~A. Rossi, A.~F. Siu, R.~Zhang, and T.~Derr.
\newblock {Knowledge Graph Prompting for Multi-Document Question Answering}.
\newblock In {\em AAAI}, 2024.

\bibitem{WWWLZLC20}
C.~Wei, B.~Wu, S.~Wang, R.~Lou, C.~Zhan, F.~Li, and Y.~Cai.
\newblock {Analyticdb-v: A Hybrid Analytical Engine Towards Query Fusion for
  Structured and Unstructured Data}.
\newblock {\em PVLDB}, 13(12):3152--3165, 2020.

\bibitem{WeikumDRS21}
G.~Weikum, X.~L. Dong, S.~Razniewski, and F.~M. Suchanek.
\newblock {Machine Knowledge: Creation and Curation of Comprehensive Knowledge
  Bases}.
\newblock {\em Found. Trends Databases}, 10(2-4):108--490, 2021.

\bibitem{WHBQRXS23}
Y.~Wu, N.~Hu, S.~Bi, G.~Qi, J.~Ren, A.~Xie, and W.~Song.
\newblock {Retrieve-Rewrite-Answer: {A} KG-to-Text Enhanced LLMs Framework for
  Knowledge Graph Question Answering}.
\newblock {\em CoRR}, abs/2309.11206, 2023.

\bibitem{WuMCJLZCY21}
Y.~Wu, K.~Ma, Z.~Cai, T.~Jin, B.~Li, C.~Zheng, J.~Cheng, and F.~Yu.
\newblock {Seastar: Vertex-centric Programming for Graph Neural Networks}.
\newblock In {\em EuroSys}, 2021.

\bibitem{wu2020comprehensive}
Z.~Wu, S.~Pan, F.~Chen, G.~Long, C.~Zhang, and S.~Y. Philip.
\newblock {A Comprehensive Survey on Graph Neural Networks}.
\newblock {\em {IEEE} Trans. Neural Networks Learn. Syst.}, 2020.

\bibitem{WuPCLZY21}
Z.~Wu, S.~Pan, F.~Chen, G.~Long, C.~Zhang, and P.~S. Yu.
\newblock {A Comprehensive Survey on Graph Neural Networks}.
\newblock {\em {IEEE} Trans. Neural Networks Learn. Syst.}, 32(1):4--24, 2021.

\bibitem{00010YAWP021}
F.~Xia, K.~Sun, S.~Yu, A.~Aziz, L.~Wan, S.~Pan, and H.~Liu.
\newblock {Graph Learning: A Survey}.
\newblock {\em {IEEE} Trans. Artif. Intell.}, 2(2):109--127, 2021.

\bibitem{XLXTWLL23}
J.~Xia, H.~Lin, Y.~Xu, C.~Tan, L.~Wu, S.~Li, and S.~Z. Li.
\newblock {GNN Cleaner: Label Cleaner for Graph Structured Data}.
\newblock {\em {IEEE} Trans. Knowl. Data Eng.}, 2023.

\bibitem{XuHLJ19}
K.~Xu, W.~Hu, J.~Leskovec, and S.~Jegelka.
\newblock {How Powerful are Graph Neural Networks?}
\newblock In {\em ICLR}, 2019.

\bibitem{XuCGWDWL24}
Z.~Xu, M.~J. Cruz, M.~Guevara, T.~Wang, M.~Deshpande, X.~Wang, and Z.~Li.
\newblock {Retrieval-Augmented Generation with Knowledge Graphs for Customer
  Service Question Answering}.
\newblock In {\em SIGIR}, 2024.

\bibitem{YangSX0LB20}
R.~Yang, J.~Shi, X.~Xiao, Y.~Yang, J.~Liu, and S.~S. Bhowmick.
\newblock {Scaling Attributed Network Embedding to Massive Graphs}.
\newblock {\em PVLDB}, 14(1):37--49, 2020.

\bibitem{YasunagaBR0MLL22}
M.~Yasunaga, A.~Bosselut, H.~Ren, X.~Zhang, C.~D. Manning, P.~Liang, and
  J.~Leskovec.
\newblock {Deep Bidirectional Language-Knowledge Graph Pretraining}.
\newblock In {\em NeurIPS}, 2022.

\bibitem{YasunagaRBLL21}
M.~Yasunaga, H.~Ren, A.~Bosselut, P.~Liang, and J.~Leskovec.
\newblock {QA-GNN: Reasoning with Language Models and Knowledge Graphs for
  Question Answering}.
\newblock In {\em NAACL-HLT}, 2021.

\bibitem{YingBYZL19}
Z.~Ying, D.~Bourgeois, J.~You, M.~Zitnik, and J.~Leskovec.
\newblock {GNNExplainer: Generating Explanations for Graph Neural Networks}.
\newblock In {\em NeurIPS}, 2019.

\bibitem{Yu0Y022}
D.~Yu, C.~Zhu, Y.~Yang, and M.~Zeng.
\newblock {JAKET: Joint Pre-training of Knowledge Graph and Language
  Understanding}.
\newblock In {\em AAAI}, 2022.

\bibitem{yu2022gpu}
Y.~Yu, D.~Wen, Y.~Zhang, L.~Qin, W.~Zhang, and X.~Lin.
\newblock {GPU-accelerated Proximity Graph Approximate Nearest Neighbor Search
  and Construction}.
\newblock In {\em ICDE}, 2022.

\bibitem{YTHJ20}
H.~Yuan, J.~Tang, X.~Hu, and S.~Ji.
\newblock {XGNN: Towards Model-level Explanations of Graph Neural Networks}.
\newblock In {\em KDD}, 2020.

\bibitem{yuan2022explainability}
H.~Yuan, H.~Yu, S.~Gui, and S.~Ji.
\newblock {Explainability in Graph Neural Networks: A Taxonomic Survey}.
\newblock {\em TPAMI}, 45(5):5782--5799, 2023.

\bibitem{YuanYWLJ21}
H.~Yuan, H.~Yu, J.~Wang, K.~Li, and S.~Ji.
\newblock {On Explainability of Graph Neural Networks via Subgraph
  Explorations}.
\newblock In {\em ICML}, 2021.

\bibitem{YuanWX17}
S.~Yuan, X.~Wu, and Y.~Xiang.
\newblock {SNE: Signed Network Embedding}.
\newblock In {\em PAKDD}, 2017.

\bibitem{yue2023routing}
Q.~Yue, X.~Xu, Y.~Wang, Y.~Tao, and X.~Luo.
\newblock {Routing-Guided Learned Product Quantization for Graph-Based
  Approximate Nearest Neighbor Search}.
\newblock In {\em ICDE}, 2024.

\bibitem{zhang2023vbase}
Q.~Zhang, S.~Xu, Q.~Chen, G.~Sui, J.~Xie, Z.~Cai, Y.~Chen, Y.~He, Y.~Yang,
  F.~Yang, et~al.
\newblock {VBASE: Unifying Online Vector Similarity Search and Relational
  Queries via Relaxed Monotonicity}.
\newblock In {\em OSDI}, 2023.

\bibitem{ZBYRLML22}
X.~Zhang, A.~Bosselut, M.~Yasunaga, H.~Ren, P.~Liang, C.~D. Manning, and
  J.~Leskovec.
\newblock {GreaseLM: Graph REASoning Enhanced Language Models for Question
  Answering}.
\newblock In {\em ICLR}, 2022.

\bibitem{ZhangZSKK22}
Y.~Zhang, H.~Zhu, Z.~Song, P.~Koniusz, and I.~King.
\newblock {COSTA: Covariance-Preserving Feature Augmentation for Graph
  Contrastive Learning}.
\newblock In {\em KDD}, 2022.

\bibitem{ZhangCZ22}
Z.~Zhang, P.~Cui, and W.~Zhu.
\newblock {Deep Learning on Graphs: A Survey}.
\newblock {\em {IEEE} Trans. Knowl. Data Eng.}, 34(1):249--270, 2022.

\bibitem{ZhangLWLL22}
Z.~Zhang, Q.~Liu, H.~Wang, C.~Lu, and C.~Lee.
\newblock {ProtGNN: Towards Self-Explaining Graph Neural Networks}.
\newblock In {\em AAAI}, 2022.

\bibitem{ZhaoDDKT21}
J.~Zhao, Y.~Dong, M.~Ding, E.~Kharlamov, and J.~Tang.
\newblock {Adaptive Diffusion in Graph Neural Networks}.
\newblock In {\em NeurIPS}, 2021.

\bibitem{ZYZLR21}
K.~Zhao, J.~X. Yu, H.~Zhang, Q.~Li, and Y.~Rong.
\newblock {A Learned Sketch for Subgraph Counting}.
\newblock In {\em SIGMOD}, 2021.

\bibitem{abs-2202-08871}
T.~Zhao, G.~Liu, S.~G{\"{u}}nnemann, and M.~Jiang.
\newblock {Graph Data Augmentation for Graph Machine Learning: A Survey}.
\newblock {\em {IEEE} Data Eng. Bull.}, 46(2):140--165, 2023.

\bibitem{0003LNW0S21}
T.~Zhao, Y.~Liu, L.~Neves, O.~J. Woodford, M.~Jiang, and N.~Shah.
\newblock {Data Augmentation for Graph Neural Networks}.
\newblock In {\em AAAI}, 2021.

\bibitem{ZhaoTZJ0SASYJ22}
T.~Zhao, X.~Tang, D.~Zhang, H.~Jiang, N.~Rao, Y.~Song, P.~Agrawal, K.~Subbian,
  B.~Yin, and M.~Jiang.
\newblock {AutoGDA: Automated Graph Data Augmentation for Node Classification}.
\newblock In {\em Learning on Graphs Conference}, 2022.

\bibitem{ZhaoZW21}
T.~Zhao, X.~Zhang, and S.~Wang.
\newblock {GraphSMOTE: Imbalanced Node Classification on Graphs with Graph
  Neural Networks}.
\newblock In {\em WSDM}, 2021.

\bibitem{zhao2020song}
W.~Zhao, S.~Tan, and P.~Li.
\newblock {Song: Approximate Nearest Neighbor Search on GPU}.
\newblock In {\em ICDE}, 2020.

\bibitem{zhao2023towards}
X.~Zhao, Y.~Tian, K.~Huang, B.~Zheng, and X.~Zhou.
\newblock {Towards Efficient Index Construction and Approximate Nearest
  Neighbor Search in High-Dimensional Spaces}.
\newblock {\em PVLDB}, 16(8):1979--1991, 2023.

\bibitem{ZhengZCSNYC020}
C.~Zheng, B.~Zong, W.~Cheng, D.~Song, J.~Ni, W.~Yu, H.~Chen, and W.~Wang.
\newblock {Robust Graph Representation Learning via Neural Sparsification}.
\newblock In {\em ICML}, 2020.

\bibitem{Zheng0WZSSGZK20}
D.~Zheng, C.~Ma, M.~Wang, J.~Zhou, Q.~Su, X.~Song, Q.~Gan, Z.~Zhang, and
  G.~Karypis.
\newblock {DistDGL: Distributed Graph Neural Network Training for Billion-Scale
  Graphs}.
\newblock In {\em IA3}, 2020.

\bibitem{abs-2309-10979}
X.~Zheng, Y.~Liu, Z.~Bao, M.~Fang, X.~Hu, A.~W. Liew, and S.~Pan.
\newblock {Towards Data-centric Graph Machine Learning: Review and Outlook}.
\newblock {\em CoRR}, abs/2309.10979, 2023.

\bibitem{ZhuZYLZALZ19}
R.~Zhu, K.~Zhao, H.~Yang, W.~Lin, C.~Zhou, B.~Ai, Y.~Li, and J.~Zhou.
\newblock Aligraph: {A} comprehensive graph neural network platform.
\newblock {\em Proc. {VLDB} Endow.}, 12(12):2094--2105, 2019.

\bibitem{0001XYLWW21}
Y.~Zhu, Y.~Xu, F.~Yu, Q.~Liu, S.~Wu, and L.~Wang.
\newblock {Graph Contrastive Learning with Adaptive Augmentation}.
\newblock In {\em WWW}, 2021.

\bibitem{ZhuXTQ19}
Z.~Zhu, S.~Xu, J.~Tang, and M.~Qu.
\newblock {GraphVite: A High-Performance CPU-GPU Hybrid System for Node
  Embedding}.
\newblock In {\em WWW}, 2019.

\bibitem{ZQZLD4SERF}
C.~Zuo, M.~Qiao, W.~Zhou, F.~Li, and D.~Deng.
\newblock {SeRF: Segment Graph for Range-Filtering Approximate Nearest Neighbor
  Search}.
\newblock In {\em SIGMOD}, 2024.

\end{thebibliography}
\end{small}

\end{document}